\documentclass[preprint]{aastex}

\shorttitle{Structure in the Halo}
\shortauthors{Yanny et al.}

\begin{document}

\title{Identification of A-colored Stars and Structure in the Halo of
the Milky Way from SDSS Commissioning Data\altaffilmark{1}}

\author{Brian Yanny\altaffilmark{\ref{FNAL},\ref{EFA}},
Heidi Jo Newberg\altaffilmark{\ref{RPI},\ref{EFA}},
Steve Kent\altaffilmark{\ref{FNAL}},
Sally A. Laurent-Muehleisen\altaffilmark{\ref{LLNL}},
Jeffrey R. Pier\altaffilmark{\ref{USNOFS}},
Gordon T. Richards\altaffilmark{\ref{UC}},
Chris Stoughton\altaffilmark{\ref{FNAL}},
John E. Anderson, Jr.\altaffilmark{\ref{FNAL}},
James Annis\altaffilmark{\ref{FNAL}},
J. Brinkmann\altaffilmark{\ref{APO}}, 
Bing Chen\altaffilmark{\ref{JHU}},
Istv\'an Csabai\altaffilmark{\ref{HUNGRY}},
Mamoru Doi\altaffilmark{\ref{UTokyo}},
Masataka Fukugita\altaffilmark{\ref{CosmicRay},\ref{IAS}},
G. S. Hennessy\altaffilmark{\ref{USNO}},
\v{Z}eljko Ivezi\'{c}\altaffilmark{\ref{PU}},
G. R. Knapp\altaffilmark{\ref{PU}},
Robert Lupton\altaffilmark{\ref{PU}},
Jeffrey A. Munn\altaffilmark{\ref{USNOFS}},
Thomas Nash\altaffilmark{\ref{FNAL}},
Constance M. Rockosi\altaffilmark{\ref{UC}},
Donald P. Schneider\altaffilmark{\ref{PSU}}, 
J. Allyn Smith\altaffilmark{\ref{UM}},
Donald G. York\altaffilmark{\ref{UC}}
}

\altaffiltext{1}{Based on observations obtained with the Sloan Digital Sky Survey, and with the
Apache Point Observatory 3.5-meter telescope, which is owned and
operated by the Astrophysical Research Consortium}

\altaffiltext{2}{Fermi Accelerator National Laboratory, P.O. Box 500, Batavia,
IL 60510\label{FNAL}}

\altaffiltext{3}{Dept. of Physics, Applied Physics and Astronomy, Rensselaer
Polytechnic Institute Troy, NY 12180\label{RPI}}

\altaffiltext{4}{LLNL/IGPP 7000 East Ave., L-413, Livermore, CA 94550\label{LLNL}}

\altaffiltext{5}{U.S. Naval Observatory, Flagstaff Station, P.O. Box
1149, Flagstaff, AZ 86002-1149\label{USNOFS}}

\altaffiltext{6}{University of Chicago, Astronomy \& Astrophysics
Center, 5640 S. Ellis Ave., Chicago, IL 60637\label{UC}}

\altaffiltext{7}{Apache Point Observatory, P.O. Box 59, Sunspot, NM 88349-0059\label{APO}}

\altaffiltext{8}{Department of Physics and Astronomy, The Johns Hopkins
University, 3701 San Martin Drive, Baltimore, MD 21218, USA\label{JHU}}

\altaffiltext{9}{Department of Physics of Complex Systems, E\"otv\"os
University, P\'azm\'any P\'eter s\'et\'any 1/A, Budapest, H-1117, Hungary\label{HUNGRY}}

\altaffiltext{10}{Department of Astronomy and Research Center for the Early Universe,
        School of Science, University of Tokyo, Hongo, Bunkyo,
Tokyo, 113-0033 Japan
\label{UTokyo}}

\altaffiltext{11}{Institute for Cosmic Ray Research, University of
Tokyo, Midori, Tanashi, Tokyo 188-8502, Japan
\label{CosmicRay}}

\altaffiltext{12}{Institute for Advanced Study, Olden Lane,
Princeton, NJ 08540
\label{IAS}}

\altaffiltext{13}{U.S. Naval Observatory, 3450 Massachusetts Ave., NW,
Washington, DC 20392-5420\label{USNO}}

\altaffiltext{14}{Princeton University Observatory, Princeton, NJ 08544\label{PU}}

\altaffiltext{15}{Department of Astronomy and Astrophysics, The Pennsylvania
State University, University Park, PA 16802\label{PSU}}

\altaffiltext{16}{University of Michigan, Department of Physics,
        500 East University, Ann Arbor, MI 48109
\label{UM}}

\altaffiltext{17}{Equal first authors\label{EFA}}

\begin{abstract}

A sample of 4208 objects with magnitude $15 < g^* < 22$ and colors of
main sequence A stars have been selected from 370 square degrees of
Sloan Digital Sky Survey (SDSS) commissioning observations.  The data
is from two long, narrow stripes, each with an opening angle of greater 
than 60$^\circ$, at Galactic latitudes $36^\circ < \vert b\vert < 63^\circ$
on the celestial equator.  Relative photometric
calibrations good to 2\% and consistent absolute photometry allows
this uniform sample to be treated statistically over the large area.
An examination of the sample's distribution shows that these stars
trace considerable substructure in the halo.  Large overdensities of
A-colored stars in the North at $\rm (l,b,R) = (350,50,46\> kpc)$ and
in the South at $\rm (157,-58, 33\> kpc)$ and extending over tens of
degrees are present in the halo of the Milky Way.  \citet{ietal00} has
detected the northern structure from a sample of RR Lyraes in the
SDSS.

Using photometry to separate the stars by surface gravity, both
structures are shown to contain a sequence of low surface gravity
stars consistent with identification as a blue horizontal branch
(BHB).  Both structures also contain a population of high surface
gravity stars two magnitudes fainter than the BHB stars, consistent
with their identification as blue stragglers (BSs).  The majority of
the high surface gravity stars in the Galactic halo may be blue
straggler stars like these.  A population of F stars associated with
the A star excess in the southern structure is detected (the F stars
in the northern structure at 46 kpc would be too faint for the SDSS to
detect).  From the numbers of detected BHB stars, lower limits to the
implied mass of the structures are $6\times 10^6 M_\odot$ and $2\times
10^6 M_\odot$, though one does not yet know the full spatial extent of the
structures.  The fact that two such large clumps have been detected in
a survey of only 1\% of the sky indicates that such structures are not
uncommon in the halo.

Simple spheroidal parameters are fit to a complete sample of the
remaining unclumped BHB stars and yield (at $r<40$ kpc) a fit to a
halo distribution with flattening ($c/a = 0.65\pm 0.2$) and a density
falloff exponent of $\alpha = -3.2 \pm 0.3$.

\end{abstract}

\keywords{Galaxy: structure --- Galaxy: halo --- stars: early-type --- blue stragglers --- horizontal branch}

\section{INTRODUCTION\label{intro}}

\subsection{Halo structure\label{background}}

Observations of distinct populations of stars in the Milky Way Galaxy
(young, metal-rich stars in the disk and older, metal-poor stars in the halo)
led \citet[ELS]{els62} to propose that the Galactic halo
formed before the
disk, during the rapid collapse of a gas cloud.
In this picture, the stars in the halo are approximately the same age,
with abundances which increase towards the center of the Galaxy.  The
model predicts a relatively smooth form and shape for the halo.
Simple models, for instance, parameterize the distribution
of matter in the Galactic halo as a flattened spheroid,
where the flattening is given by an axial ratio $c/a \leq 1$ (and $b/a
\le 1$ if triaxiality is considered), with a power-law density profile
$\rho(r) \sim r^\alpha$.

As the positions and motions of luminous objects in the Galactic halo
in principle provide tracers of mass within the halo potential, study
of these objects is important for understanding the
distribution of dark matter as well as providing clues to the
formation of the Galaxy.

Model parameters for this simple description of the halo have been
obtained by empirical fits to the distributions of such astronomical
objects as Galactic globular clusters, K dwarfs, and A-colored stars.
Here the term ``A-colored stars'' includes all types of stars,
including blue horizontal branch (BHB) and blue straggler (BS) stars,
which have colors and spectra similar to those of the bright, nearby,
main sequence A stars on which the spectral classification was
defined.  The distribution of Galactic globular clusters yields a fit
to the structure parameters of $c/a \sim 1$ and $\alpha \sim -3.5$
\citep{h76,z85}.  \citet{s85} used RR Lyrae stars to find
$\alpha \sim -3$ out to distances of 25 kpc, with a steeper decline
thereafter.  \citet{ietal00} also measured a distribution of RR Lyrae
stars with $\alpha \sim -3$ to at least 50 kpc.  Hubble Space
Telescope counts of K dwarfs by \citet{gfb98} yield $c/a = 0.8 \pm
0.1$, and $\alpha = -3.06 \pm 0.22$.

It is of course not required that $c/a$ remain constant with
Galactocentric radius.  \citet{psb91} counted BHB stars and found
evidence that $c/a$ increases from 0.5 to 1 as one moves out to $R >
20$ kpc with $\alpha = -3.5$.  Using this changing $c/a$, \citet{wm96}
find $\alpha = -3.53\pm 0.077$ for RR Lyraes. \citet{sa98} find a flat
$c/a$, but less evidence for changing flattening with Galactocentric
radius.  As part of an extensive kinematic analysis of several hundred
nearby halo stars with [Fe/H] $\le -1.8$, \citet{cb00} find that the
axial ratio increases from 0.65 at $R < 15$ kpc to 1.0 at $R \sim 20$
kpc.  Most of these studies use a few hundred objects at most on which
to base their results, and only a few, such as \cite{cb00}, have
kinematic information.

In contrast to the ELS model, recent research suggests that some or
all of the halo may have been formed through accretion of smaller
dwarf galaxies \citep[see][and references therein]{j99,m99}.  One
example of a dwarf galaxy thought to be in the process of accreting
has been discussed by \citet{igi95}.  As these smaller galaxies fall
into the Milky Way, they may be tidally disrupted and break up into
long streams of stars which can persist for 1-10 Gyr
\citep{ll95,j98,hwzz99}.


\subsection{The nature of A-colored stars in the Galactic halo\label{nature}}

A-colored stars at faint magnitudes are usually assumed to be halo
population (Pop II) stars, and hence BHB stars.  However, it has become
clear that many distant blue stellar objects have hydrogen line widths
indicative of main sequence gravity A stars rather than the less
massive BHB stars \citep{rhs81,scc89,bpsdg92,pbs94,wbsplfrc99}.

Why can't all of these high-gravity A-colored stars be normal
main-sequence stars?  The well-known arguments against this are
two-fold: First, except for dwarf irregular companions to the Milky
Way such as the LMC and the Sextans dwarf, there is no evidence that
significant star formation has taken place in the far halo (R $>$ 30
kpc) within the last billion years, which is greater than the
main-sequence lifetime of an A star.  Second, the stars are too far
from the star forming regions in the Galactic plane to have reached
their present positions in the far halo with any reasonable assumed
velocity (it takes 300 Myr to travel 30 kpc at 100 km s$^{-1}$).

Blue straggler stars in globular clusters, which have
ages closer to 10 Gyr, are an interesting example of
high surface gravity A-colored objects which are present in an old
environment.  These high surface gravity BS stars are generally
thought to be main sequence stars which result from
mergers of, or mass transfer between, two lower mass stars
\citep{s93,sb99}.  Whatever their formation mechanism, they
should also be present at some level in the field of the Galactic halo
\citep{rr93,nh91}.

\subsection {The Sloan Digital Sky Survey\label{sdss}} 

The Sloan Digital Sky Survey (SDSS) is a large, international
collaboration set up to survey 10,000 square degrees of sky in five
optical passbands and to obtain spectra of a million galaxies, one
hundred thousand quasars, and tens of thousands of Galactic stars.  The
data are being taken with a dedicated 2.5 meter telescope located at
Apache Point Observatory (APO) near Cloudcroft, New Mexico.  The
telescope has two instruments: a CCD camera with 30 2K x 2K CCDs in
the focal plane, and two 320-fiber double spectrographs.  The imaging
data are tied to a network of brighter astrometric standards (which
would be saturated in the main imaging data) through a set of
twenty-two smaller CCDs in the focal plane of the imaging camera.  
A 0.5 m telescope at APO will be
used to tie the imaging data to brighter photometric standards.  

The USNO 1m telescope at Flagstaff, AZ, has been used to define
the photometric system, defining SDSS standard stars in $u', g', r',
i',$ and $z'$ (Smith et al., in preparation).
Since
the precise calibration for the SDSS filter system is still in
progress, magnitudes in this paper are quoted in the $u^* g^* r^* i^*
z^* $ system, which approximates the final SDSS system.  These systems
differ absolutely (with negligible color terms) by only a few percent
in $g^* r^* i^* z^*$, and no more than 10\% in $u^*$.  See
\citet{yetal00}, \citet{figdss96} and \citet{getal98} for further
information on the SDSS, the filter system and the imaging camera.

The SDSS has imaged a large fraction of the celestial equator ($\rm
-1.25^\circ < \delta < 1.25^\circ$) as part of telescope and imaging
camera commissioning.  Catalogs produced from nearly 400 square
degrees of these data contain thousands of A-colored stars suitable
for probing the halo, with absolute photometry good to $\sim5$\% for
objects with $g^* < 20$. For reference, blue stars with $0 < B-V <
0.2$ (including A stars), have a SDSS $g^*$ magnitude approximately
equal to their Johnson $V$ magnitude.

\section {THE SAMPLE OF A-COLORED STARS\label{sample}} 
\subsection{Photometric Observations and Data Samples\label{photodata}}

The SDSS camera contains six columns of CCDs, each of which records
data for a 0.23 degree wide strip of the sky.  The CCDs are
continuously read out as the camera scans a great circle across the
sky at the sidereal rate, so the length of an imaged strip grows at 15
degrees per hour.  The effective exposure time in each of the 5
filters is 54 seconds.  A ``run'' generates six parallel continuous
strips of sky with a gap of 0.2 degrees between columns.  A second
``run,'' offset slightly from the first, is obtained to fill in the
sky between the first six scan lines and fully image a 2.5 degree wide
``stripe'' on the sky.

Figure \ref{fg1} shows $u^*-g^*$ vs. $g^*-r^*$ for all stellar
objects with $g^* < 21$ in 13.3 square degrees of one column of SDSS
data.  Objects in different magnitude ranges are plotted in different
colors.  As discussed by \citet{nrrf99}, the position of the blue end
of the Galactic stellar locus shifts as a function of magnitude due to
the sampling of populations of stars with different characteristic
metalicities at different scale heights in the disk.  The temperatures
as a function of color can be estimated by comparison with
\citet{lnrrs98}.  The stars in the A box in the diagram are expected to
contain nearly exclusively hot stars with a strong Balmer jump -- BHB
stars, BSs, and main sequence stars with spectral type A.  These bluer
objects do not shift by large amounts in color with magnitude.  Boxes
for quasar candidates, F, and G type stars are also indicated and will
be discussed below.

All of the magnitudes used in this paper are point spread function
(PSF) fitted magnitudes for stellar objects, generated by summing
the PSF-weighted flux for each object (Lupton et al., in preparation).
Saturation sets in for A-colored objects brighter than $g^* \sim 15$.
All magnitudes in this paper use the modified flux designation for
faint magnitudes described in \citet{lgs99}, though for the range of
magnitudes relevant here, $15 < g^* < 22.5$, the distinction is
negligible.

The boundaries of the A-colored star box were chosen to provide a
sample of A stars, BHBs and BSs, free from contamination from other
populations of objects even to magnitudes as faint as $g^* \sim 21$.  The
choice of a $g^*-r^* = 0 $ red limit was conservatively chosen in order to
ensure that the Gaussian tail of color measurement errors of the
population of the very large numbers of F stars did not begin to
scatter into the A star box at faint magnitudes.

All magnitudes and colors in the present work have been corrected for
reddening using the maps of \citet{sfd98}.  Since all of the A-colored
objects sampled here are at distances $\rm d > > 1 \>kpc$ from the plane
of the Galactic disk, where most reddening-producing dust occurs, the
full reddening correction of \citet{sfd98} is applied to all objects.

For this analysis, two samples of A-colored stars are chosen from two
stripes of data.  The northern sample (north of the Galactic plane on
the celestial equator) is a filled stripe from SDSS runs 752 and 756,
obtained 1999 March 21 and 22.  The stripe is 87 degrees long ($\rm
145.5^\circ <\alpha<232.5^\circ$ [J2000], and 2.52 degrees wide ($\rm
-1.26^\circ < \delta < 1.26^\circ$), with 219 square degrees of
data. Data in the region $\rm 215.5^\circ <\alpha< 216^\circ$ from run
752 was excluded, since the seeing increased sharply to $3''$ in this
region (compared to typical seeing of $1.3''$ for the rest of the
scan).  In Galactic coordinates, the northern sample runs in a long
arc from [l,b] = [236,37] up to [303, 63] and then gradually back to
[4, 43].

The northern sample contains 3126 point sources (extended sources are
rejected) in the runs 752 and 756 which have dereddened colors in the
box in Figure \ref{fg1} ($0.8 < u^*-g^* < 1.5$ and $-0.3 < g^*-r^* <
0.0$), and which have $u^*$ errors smaller than 30\% (corresponding to
a limit of $g^* \sim 22.5$).

The southern sample (south of the Galactic plane on the celestial
equator) is a filled stripe from SDSS runs 94 and 125, obtained 1998
September 19 and 25.  The stripe is 60 degrees long ($\rm 353^\circ
<\alpha< 53^\circ$) and 2.5 degrees wide ($\rm -1.24^\circ < \delta <
1.27^\circ$), with 151 square degrees of data.  The southern sample
contains 1082 stars, to $g^* \sim 22$, with the same selection
criteria as the northern sample. The southern sample runs from [l,b] =
[85,-57] through [125,-63] and then back to [184,-43].

The SDSS has imaged the majority of these equatorial regions twice.
SDSS runs 259, 273 and 745 (obtained 1998 November 17, 19 and 1999
March 20, respectively), nearly completely overlap runs 94, 125 and
756, and are used to verify the findings in the primary runs.

The list of A-colored stars from runs 745 and 756 were matched
(without the requirement that the error in $u^*$ be smaller than
$30\%$), based on position (within one arcsecond, well in excess of
typical one sigma astrometric errors of 100 mas).  Additionally,
objects from these overlapping scans were matched without regard to
color in order to estimate completeness, detection limits and
reproducibility of the data.  The upper curve in Figure \ref{fg2}
shows the fraction of objects matched as a function of $g^*$ for all
stellar sources.  The lower and noisier curve is that for matched
objects within the A-colored box only.  The maximum level of
$\sim80$\% matched in the lower curve is primarily due to objects in
one sample or the other scattering out of the box.  One can estimate
this fraction by noting that most of the objects in the box are on the
red side, giving the box an effective width of 0.15 mag in $g^*-r^*$.
When combined with typical color errors of 2\% for two measurements,
one derives an expected fraction $f = 1- 0.02*\sqrt{2}/0.15 = 81\%$.
The matched fraction decreases rapidly for $g^* > 20.0$.  Here, the
requirement that the object fall within the box in $u^*-g^*$ becomes
the determining factor.  From Figure \ref{fg2}, one sees that the
A-colored star sample is uniform (in the sense that it samples the
same population of stars) to $g^* \sim 20$ and has
detections to about $g^* \sim 22.5$ (50\% limit) in the northern
stripe.  After $g^* \sim 20$ the population of stars includes an
increasing fraction of stars slightly redward of the box whose
photometric errors have scattered them into the box.  The fact that
the upper curve doesn't start to drop until $g^* \sim 22.5$ indicates
that the uniform sample is not affected by the detection limit, and is
in that sense complete.  A related procedure in the south yields a
uniform sample to $g^* = 19.5$ and detections to $g^*
\sim 22$.

\subsection{Distribution of the A-colored stars\label{distrib}}

Horizontal branch stars, especially those in a restricted color range,
are approximately standard candles.  A BHB absolute magnitude of
$M_{\rm g^*} = +0.7$ can be used to infer approximate heliocentric
distances from magnitudes for these objects.  This absolute magnitude
was calculated by converting the V absolute magnitude of halo RR Lyrae
stars from \citet{lhhkh96} into the $g^*$ filter, using the
transformations from \citet{figdss96} for the colors of RR Lyrae
stars.  Since the $g^*$ magnitudes of RR Lyrae stars in Pal 5 (Figure
\ref{fg7}) are similar to the BHB magnitudes in Pal 5, the assumption
has been made throughout this paper that the absolute magnitudes of
BHBs are the same.  This approximation is dependent on the blue
response of the filter used.  Note, for example, that \citet{psb91},
find that for objects with colors at the blue end of the A star
selection box, BHBs are about 0.9 magnitudes fainter in V than the RR
Lyrae magnitude \citep[this corresponds to a difference of 0.7
magnitudes in $g^*$, given the transformations in][]{figdss96}.

Figure \ref{fg3} presents a wedge diagram with Right Ascension and
$g^*$ magnitude for all 4208 A-colored stars from both
North and South samples.  Significant structures are immediately
apparent.  The clump at [l,b] = [242,43], or $(\alpha,\delta)=(153,0)$,
and $20.50 < g^* < 20.55$ is the northern half of the Sextans dwarf
irregular galaxy \citep{vpw98} at $d = 70$ kpc (the southern half of
Sextans is at $\rm \delta < -1.25^\circ$ and is not covered in this
data set).  One also notes two dramatic arcs of structure at $g^* \sim
19$ and $g^* \sim 21$ in the North ($\rm 195^\circ <\alpha<
230^\circ$), and similar, though less pronounced, structure at $g^* = 18$ and
$g^* \sim 20$ in the South ($\rm 20^\circ < RA < 45^\circ$).

The large structure in the north was detected by \citet{ietal00}
in a search for variable RR Lyrae stars in the overlapping SDSS runs
745 and 756 taken two days apart.  It contains a remarkable
population of $\sim 80$ RR Lyraes clustered in the same way as the
bluer A-colored stars in the northern sample.

There are numerous smaller apparent concentrations throughout this
diagram.  The Galactic globular cluster, Pal 5, for example, has its
horizontal branch at $g^* = 17.3$, $\rm \alpha=229^\circ$ in the
North.  An enlargement of the three degrees of arc in the Pal 5 area of
sky is shown to the right in Figure \ref{fg3}.  Along with the stars
of Pal 5's BHB at $g^* = 17.3$, the narrow sequence of Pal 5 BSs (see
\S\ref{pal5example}) is clearly visible from $18.5 < g^* < 20$, also
at $\rm \alpha=229^\circ$.

\subsection {Verification of the Halo Substructure\label{checks}}

A number of straight-forward checks are performed to verify that
the large structures in Figure \ref{fg3} are not artifacts.

In order to assess whether these structures could be due to selection
effects, low-redshift quasar candidates are color selected from the
SDSS survey from a color box with boundaries $-0.05 < u^*-g^* < 0.15$,
$-0.1 < g^*-r^* < 0.0$ (the small Q box in Figure \ref{fg1}) and plotted in
Figure \ref{fg4}.  Figure \ref{fg4} shows no significant magnitude
(and thus implied spatial) structure to $g^* \sim 22$, except for the
increasing density of quasar candidates to fainter magnitudes.  While
not every object in this color box is a quasar, other hot blue sources
which contaminate the low redshift quasar population, such as white
dwarfs and narrow emission line compact galaxies, also have population
distributions which are expected to be isotropic, showing no
structural signature of the Galactic disk, bulge or halo
\citep{fetal00}.  The uniformity of the distribution in Figure
\ref{fg4} demonstrates the uniformity of the photometry and
calibration in the SDSS samples to magnitude limits of $g^* \sim 22$
for objects with colors similar to A stars.  Fainter than $g^* \sim
22$, the errors in the $u^*$ magnitudes of F stars become large enough
that some of them scatter into the quasar color box.  The distribution
of F stars in the northern and southern sample areas is discussed in
\S\ref{fgstars}.

The photometric calibration within an individual stripe is accurate to
a few percent due to the continuous drift scanning-method of data
collection and photometric observing conditions.  Relative photometric
accuracy of objects detected in multiple scans is good to $\sim 2\%$
for blue objects to $g^* \sim 20$ in the $g^* r^* i^*$ and $z^*$
filters.  The absolute calibration, and the calibration between runs
observed in different seasons, relies upon calibration patches
obtained with a small photometric telescope, and is not yet final.

To verify that the absolute calibration is essentially identical in
the northern and southern stripes, the density of quasar candidates
(chosen from the larger Q box in Figure \ref{fg1} to increase
statistics) as a function of magnitude for the North and South is
plotted in Figure \ref{fg5}.  The excellent agreement between the two
data sets to $g^* \sim 21.5$ is further evidence that the photometric
calibration is uniform across the sky.  One notes from the earlier
falloff for southern quasar candidates in Figure \ref{fg5} that the
southern sample completeness limit is about 0.5 magnitudes brighter
than that of the northern sample.  As the objects are selected in the
bluest SDSS filters, and since the $u'$ filter is the least sensitive
filter, completeness limits and the magnitude limit for detections of
objects in the present work is determined primarily by where the error
in the $u^*$ band flux for objects becomes large.  Note that although
$g^*-r^*$ colors for quasars and A stars are similar, the $u^*-g^*$
colors of quasars are bluer than A stars by about one magnitude.
Therefore, one expects to see a drop in A-star counts one magnitude
brighter than the drop in quasar counts, as is verified by comparing
the upper curve in Figure \ref{fg2} to the northern points in Figure
\ref{fg5}.

The duplicate equatorial runs 745 and 273 were used to generate a plot
similar to that shown in Figure \ref{fg3}.  The structures noted in
\S\ref{distrib} are also clearly visible in Figure \ref{fg6}, though
there are fewer stars since only half of a filled stripe is plotted.
Note that run 745 doesn't cover Sextans in RA, but does extend beyond
the end of the northern structure, and thus more clearly defines its
full length than in Figure \ref{fg3}.

\section {SPECTROSCOPIC FOLLOW-UP\label{spectro}}

\subsection{Observations}

While the SDSS continues commissioning of its spectrographs, which will
obtain thousands of spectra of type A and BHB stars in addition to
its main galaxy and quasar redshift survey, a small number of
representative spectra have been obtained with other telescopes 
to investigate the nature of the selected A-colored objects.

The observations produced spectra for a sample of 6 A-colored
stars selected from the SDSS samples, along with spectra of 5 A-colored
reference stars with known surface gravities.

The Lick Observatory 3 m telescope was used to obtain reference stars,
including CS 29516-0011, a BHB identified as a field horizontal
branch star by \citet{wbsplfrc99}; WD 0148+467, a DA white dwarf
identified by \citet{gj79}; and HD 13433, a known A0V star classified
by \citet{hs88}, all with $8 < V < 12$.  These spectra
were obtained with exposures between 5 s and 60 s on the night of 1998
Dec 19 at a resolution of about 6\AA, obtaining good S/N ratios of
about 50:1 in the continua.

Figure \ref{fg7} shows the color-magnitude diagram of Pal 5, generated
from reddening-corrected SDSS data.  Stars of all colors within $\sim
9$ arcminutes of the center of the globular cluster Pal 5 were
selected; note that there is significant field contamination here at
[l,b]=[1,46].  The vertical lines show the $g^*-r^*$ limits of the
A-colored star box.  A 5400 s exposure of one blue straggler and a
1800 s spectrum of a BHB star (marked with circled squares in
Fig. \ref{fg7}) were taken at 3\AA\ resolution with the dual imaging
spectrograph (DIS) on the ARC 3.5 meter telescope at Apache Point
Observatory on the night of 1999 May 11.  Proper motion studies
confirm that both objects are members of the Pal 5 globular cluster
(K. Cudworth, private communication).  The resolution of these spectra
is approximately 3\AA\ at a S/N $\sim 30$ for the brighter object and
15 for the fainter.  Wavelength calibrations
were obtained and standard flats and reductions were applied to all
spectral data.

Six additional 30 min spectra of $17 < g^* < 18$ A-colored stars
selected from the SDSS were observed on the nights of 1999 May 8, 10,
and 11 with the DIS at 3\AA\ resolution and S/N of $\sim 20$.  The six
spectra all confirmed that the A-colored star sample indeed consists
entirely of blue stars with strong Balmer lines, and is not
contaminated by significant numbers of white dwarfs, quasars or other
extragalactic objects such as compact E+A galaxies.

\subsection{Spectroscopic Surface Gravity Indicator}

The primary diagnostic used to distinguish BHBs from A
main sequence (or blue straggler) stars is the widths of the Balmer
line profiles as indicators of stellar surface gravity 
\citep[see][and references therein]{p83}.  Main sequence 
A stars ($M \sim 2.5 M_{\odot}, R \sim 2 R_{\odot}$) 
have $\rm 3.5 < log \> g < 5$, while horizontal
branch stars, which have much smaller mass ($M \sim
0.7 M_{\odot}$) and larger radii $\sim 3 R_{\odot}$, have $\rm 2.5 < log
\> g < 3.2$.  This difference in surface gravities clearly shows up
in the Stark pressure-broadened Balmer line profiles.  White dwarfs, which
have much higher surface gravities -- typically $\rm log \> g > 7.5$,
are easily distinguished from main sequence A stars.  Since white dwarf
$u^*-g^*$ colors are significantly bluer than those of A and BHB stars, they
are not present in the A-star selection box in significant numbers.
None of the spectra of A-colored stars presented here is consistent
with a white dwarf identification.

The standard indicator which separates main sequence gravity A stars
from lower surface gravity BHB stars is the width
of the H$\delta$ line at $80\%$ of the continuum.  One typically uses
the H$\delta$ line because the A star flux is higher here than in longer
wavelength Balmer lines, and it is easier to fit the continuum at
H$\delta$ than at shorter wavelength Balmer lines.  The width at
$80\%$ of the continuum is a measurement robust to the presence of
metal lines which can apparently widen the absorption lines near the
continuum, and is insensitive to a range of instrumental resolutions,
which can change the measured depth of the line.  


IRAF was used to extract the 1D spectra, select a region of the
spectrum containing the H$\delta$ and H$\gamma$ lines, divide the
continuum by a
best fit 6$\rm th$ order Legendre polynomial, and then boxcar smooth
and wavelength shift the spectra to match the resolution and redshift
of the Lick reference spectra.  The difference in resolution between
the sets of observations is not critical, since the broad Balmer lines
typically have equivalent widths of 30\AA.  It must be noted that with
very wide Balmer lines it is difficult to uniformly fit the continuum in
such a way that the lines themselves do not affect the fit.

The results are plotted in Figures \ref{fg8}-\ref{fg9}. Figure
\ref{fg8} shows the Pal 5 stars compared to the blue reference
templates.  The BHB star in Pal 5 is a good match to the template BHB
star.  The spectrum of the Pal 5 blue straggler has much larger line
widths consistent with the template A0V star.  The six additional
representative A-colored objects were spectroscopically confirmed with
spectra shown in Figure \ref{fg9}.  For these six brighter ($17 < g^*
< 18$) SDSS northern sample targets observed from APO in Figure
\ref{fg9}, 4 have narrow widths, one has a broad width (lower left
panel), and one is apparently a metallic-lined Am star (upper left
panel).

Table 1 summarizes the spectroscopic targets, listing SDSS id (or
other id such as HD number), $\alpha$, $\delta$ , dereddened $g^*$
magnitude, $u^*-g^*$, $g^*-r^*$, $H_{\delta}(0.8)$, l, b, and
classification determined from the Balmer line widths in the spectrum.

The radial velocities of the stars were only approximately measured
(with typical error of 30 $\rm \>km\>s^{-1}$), and ranged from -120 to
$\rm +120\> km\>s^{-1}$, consistent with halo objects.  The resolution
of the observations and uncertainties in the wavelength calibrations
made it difficult to draw conclusions about clustering of this small
sample of objects from radial velocities.  Kinematic information will of
course be an important future diagnostic of SDSS A-colored stars in
continuing searches for coherent halo structures.

\section {THE NATURE OF THE A-COLORED STAR SAMPLE\label{newnature}}

\subsection{The Example of Pal 5\label{pal5example}}

As previous authors \citep{psb91,nh91,pbs94,wbsplfrc99} have noted, 
and the spectroscopy in this paper also
shows, one must be careful drawing conclusions about Galactic
structure from color-selected A-stars in the halo of the Milky Way,
since the blue star population is not homogeneous; it includes at
least two types of objects.

One gains insight into the nature of the A-star sample by examining
the appearance of the globular cluster Pal 5 in the star distribution
in Figure \ref{fg3} at $\rm\alpha= 229 ^\circ$.  Pal 5, located about
20 kpc from the Sun, contributes 5 BHB stars at $g^* \sim 17.3$, and
at least 7 BS stars at $18.5 < g^* < 20.0$ (see Figure \ref{fg7}).
On an absolute scale, the BHBs in Pal 5 have $M_{g^*} \sim 0.7$,
and the BSs are thus at absolute magnitudes of $1.9 <
M_{g^*} < 3.4$.

The large parallel arcs of A-colored stars present in Figure
\ref{fg3} are also separated by about 2 magnitudes, suggesting that
they might represent associated BHB and blue straggler stars at a
common distance.  If this assumption is correct, then one can use the
location of the BHB in $g^*$ and the absolute magnitude of BHB stars
to determine a distance to the stars in the arcs.  For example, the
northern arc, with an implied BHB magnitude of $g^* \sim 19$, is then
approximately 45 kpc from the Sun.

\subsection {Photometric determination of surface gravity\label{photog}} 
 
In order to support the identification of the stars in the 
overdensities, one would like to be able to separate all of the high 
surface gravity stars from the low surface gravity stars.  Presently 
lacking individual spectra for such a large sample, one turns to 
photometric multi-color indicators of surface gravity in the SDSS 
filters.  From a sample of 1121 A-colored stars, \citet{wbsplfrc99} 
show that they can separate (although not perfectly) high surface 
gravity stars from low 
surface gravity stars in $U-B, B-V$ color-color diagrams.  Due to the 
lack of a single-valued transformation between the $U$ and $u^*$ (they
have different behavior around the Balmer jump), a simple 
conversion between $u^*-g^*$ and $U-B$ is not well defined.  Thus, 
although one cannot precisely convert the $U-B, B-V$ color separation 
criteria to the SDSS filter system, a similar idea is applied to 
approximately separate the populations: extreme Balmer jump BHB stars 
are in general slightly bluer in $g^*-r^*$ and redder in $u^*-g^*$ 
than those with higher gravities. 
 

The other SDSS filters available ($i^*,z^*$) may be considered for
color separation of gravities.  For instance, \citet{lnrrs98} show
that in synthetic A stars derived from Kurucz model spectra
\citet{k91},
different gravities separate well in the $z'$
filter, and they derive a direction in SDSS color space that results
in maximum separation. Since $z^*$ magnitudes accurate to 2\% are
lacking, for most of the faint A star sample considered here, the
stars were separated in $u^*-g^*$ and $g^*-r^*$ only.  (The $r^*-i^*$
color contributes little to surface gravity separation for objects
with A star effective temperatures.)  Figure \ref{fg10} shows the
colors of the \citet{lnrrs98} models, along with the colors of the six
stars for which spectra exist, and the colors of the BSs and
horizontal branch stars in Pal 5.  The positions of the Pal 5 stars
plotted in Figure \ref{fg10} are circled in the Pal 5 color-magnitude
diagram in Figure \ref{fg7}.  One star which appears to be on the
horizontal branch of Pal 5 and also in the A-star box is far enough
from the center of the globular cluster that its membership is
questionable, so it was not included in Figure \ref{fg10}.  The
adopted separation between BHBs and high gravity stars (thick line in
Figure \ref{fg10}) in $u^*g^*r^*$ colors was based empirically on the
spectra and photometry obtained here, guided by \citet{lnrrs98} and
\citet{wbsplfrc99}.  The models are valid for stars with $6500 <
T_{\rm eff} < 10000 K$.  A few candidate RR Lyraes, selected as in
\citet{ietal00}, are also indicated.
 
\subsection {Subsamples of A-colored stars\label{split}} 
 
Applying the color separation in Figure \ref{fg10}, the 3126 stars in
the northern sample are split into 2041 candidate blue straggler stars
and 1085 candidate BHB stars.  The southern sample split into 674
candidate blue straggler stars and 408 candidate blue horizontal
branch stars.  The wedge diagrams for these separated stars are shown
in Figure \ref{fg11} (BHBs only) and Figure \ref{fg12} (BSs only).
These plots separate and enhance the contrast of the structures found
in Figure \ref{fg3}, supporting the identifications in \S4.1.  The
structures are interpreted as an elongated arc of BHB stars at $g^*
\sim 19$ in the North, and an arc of BS stars about 2 magnitudes
fainter in the same direction in the North.  Additionally, an arc of
BS at about $g^* \sim 20$ is apparent in the South.  Both the northern
and southern structures extend over 30 degrees or more of the sky.
The SDSS stripe is $2.5^\circ$ wide in declination and so these
large structures may extend well beyond the narrow stripes observed
here.

One also detects the candidate BHB stars associated with the candidate
BS stars in the southern structure.  Figure \ref{fg13} shows
histograms of the candidate BS and BHB stars in the RA ranges of the
northern and southern structures as a function of magnitude.  While
there are no significant structures apparent at $\alpha < 200^\circ$
in the North (left panels), the plot clearly shows an overdensity of
BHB stars at $g^* \sim 19$ in the North at $\alpha > 200^\circ$
(middle panels) and also shows the overdensity in BHBs at $g^* \sim
18$ in the South (right panels). 

The BS/BHB color-color separation is far from razor sharp.  Figure
\ref{fg11}, for example, which should show the absence of the second
fainter arc from the candidate BHB star set and leave only the
brighter arc in the north, does not do this completely.  This is
expected, since the color separation technique does not definitively
separate the two populations of stars (see Figure \ref{fg10}).  The separation
is less effective at fainter magnitudes, where the errors are larger,
and at redder colors, where the populations overlap in color-color
diagrams.

The $g^*-r^*$ color distribution of candidate BS and BHB stars in the
large structures in the North and South is histogrammed in Figure
\ref{fg14}.  In the upper panel, candidate BSs with $\alpha >
200^\circ$ and $g^* > 20$ in the Northern structure are indicated with
a light line against $g^*-r^*$ color, while candidate BHBs ($18.8 <
g^* < 19.4$) in the same area are histogrammed with a heavy line.  The
lower panel shows the Southern structure objects with $19.0 < g^* <
20.5$ BSs and $17.7 < g^* < 18.3$ BHBs.  In this color system, an A0V
main sequence star has $g^*-r^* = -0.25$, whereas a somewhat cooler
A5V star will have $g^*-r^* \sim -0.05$.  The distribution of BS stars in
Figure \ref{fg14} shows that the BS candidate stars are consistent
with intermediate (A3-A5) A star colors, and the population is not
dominated by the hottest A0 colored objects.  The BS candidate sample
is also notably consistent with the Pal 5 BS objects of Figure
\ref{fg7}, which also doesn't contain any BS stars as hot as an A0
dwarf.

One might naively expect the A colored stars in the Sextans dwarf
irregular galaxy to represent its horizontal branch.  However,
\citet{vpw98} show that this galaxy has a young very blue main
sequence; if there is a horizontal branch, it contains very many fewer
stars.  The ratio of the number counts of BS:BHB candidates is 36:19
for the Sextans dwarf.  This ratio is similar to the ratio of the
excess of BS candidates at $g^* \sim 20$ to the excess of BHB
candidates in the same magnitude range in the southern structure.  In
both cases, we assume that the candidate BHBs are misidentified high
surface gravity stars.  Thus for Sextans, nearly all the BS and BHB
candidates are simply young main sequence stars.

\subsection {Simulation of a Spheroidal Halo\label{sim}} 
 
The tight concentration of BHB magnitudes in Figure \ref{fg13} and
the existence of associated BS candidate stars are strong evidence
that the northern and southern structures are real halo substructure.
However, one must explore the possibility that these structures could
be artifacts of a smoothly varying galactic structure combined with
the magnitude limits of the SDSS samples.  One is concerned with the
possibility that the northern structure might be associated with the
bulge of the Milky Way, since it occurs at a Galactic longitude of
zero degrees.

Figure \ref{fg15} shows a simulated halo density distribution
expected for a smooth spheroidal distribution of BHB stars, assuming
$\rm M_{g^*}(BHB) = +0.7$, $\alpha = -3.0$, and $c/a = 1.0$.  One can
see from this figure that the highest density of BHBs is closer to the
Galactic plane in the equatorial stripe (indicated by the line of
nodes in Figure \ref{fg15}), than to Galactic longitude of zero.  This
is because the plane of the celestial equator passes closer to the
center of the Galaxy at the Galactic plane than at Galactic longitude
of zero.  The halo density distribution depends only on the distance 
from the Galactic center for a spherically symmetric halo population.

One must resort to an extremely unlikely model to reproduce 
an over-density such as that seen in Figure \ref{fg3} within a
smooth spheroid.  For example, assuming an unrealistic absolute
magnitude of $M_V \sim +4$ for the A-colored stars, and an extremely
prolate ($c/a = 3$) bulge, one can obtain an overdensity near $l=0$,
but it will be much more spread out in magnitude than the tight BHB
locus seen at $g^* \sim 19$.  Additionally, such an exercise can in no way
reproduce the observed overdensity in the South with the same smooth
model.  The southern structure is nearly directly opposite the
Galactic center.

\subsection { Distribution of redder Galactic stars\label{fgstars} } 

Figure \ref{fg16} shows a wedge plot, similar to Figure \ref{fg3},
but for F star candidates (objects with $ 0.6 < u^*-g^* < 1.0,\> 0.1 <
g^*-r^* < 0.3$), which are intrinsically about 3 magnitudes fainter
than BHB stars.  One sees here the familiar overdensity in the
southern sample, but the typical magnitude, $g^* \sim 21$, is about one
magnitude fainter than that of the presumed BSs of Figure \ref{fg12}.
This indicates that one may be detecting F stars associated with the
southern structure ($g^*_{BHB} = 18$).  Note that although the density
of the F star concentration appears to fall off near $\alpha=49^\circ$,
the density of QSOs in Figure \ref{fg4} also drops off in this region
at about $21^{st}$ magnitude in $g^*$, indicating that the sample is
less complete in that region.

The rest of the F star distribution is consistent with a smooth
Galactic model.  (Imagine shifting the model distribution in Figure
\ref{fg15} three magnitudes fainter.)  The halo F stars associated
with the northern structure are too faint to be unambiguously detected
with the SDSS.

A magnitude-RA wedge plot for color-selected G stars for the same SDSS
samples is presented in Figure \ref{fg17}.  It is reassuring to see
no signs of the northern or the southern structures in this wedge
plot; the G stars ($M_V \sim +5$) associated with these structures are
too faint to appear above the SDSS magnitude limits.  This population
may prove valuable for studies of Galactic structure at closer
distances than those probed by the BHB/BS stars.  A slight leak of F
stars in to the G star box may be manifesting itself as structure in
the South at $g^* \sim 21.5$.

\section {PROPERTIES OF THE LARGEST STRUCTURES\label{largeblobs}}

As horizontal branch stars have $M_{g^*} \sim 0.7$ and BS stars have
$M_{g^*} \sim 2.7\pm 1.0$, one can estimate
the heliocentric distances to the largest structures using the simple
magnitude-distance relation, $\rm m - M_{g^*} = 5 \>log(d_{pc}) - 5$.
Assuming the Sun is $8$ kpc from the Galactic center, one can then use
[l,b] for each object and calculate a Galactocentric distance for
each star.  The resulting Galactocentric distance estimates are
plotted as ($\alpha$, $\rm R_{GC}$) polar wedge diagrams in Figure
\ref{fg18} (for the BHB subsample) and Figure
\ref{fg19} (for the BS subsample). The implied distance of the large
clump of stars in the North is 46 kpc from the Galactic center, with
Galactic rectangular coordinates extending from (X,Y,Z) = (12, -13, 41)
to (26, 1, 34) centered at (X,Y,Z) = (21, -5, 38) [units are kpc, where
the sun is at (-8,0,0)].  The most significant clump in the south is
inferred to be about 33 kpc from the GC extending from (X,Y,Z) =
(-18, 9, -25) to (X,Y,Z) = (-25, 3, -22) centered at (-22, 6, -24).  The
significance of other smaller clumps seen in Figures \ref{fg18} and
\ref{fg19} is tantalizing, and merits kinematic followup to determine
if they are moving groups or streamers.  In this coordinate system,
Sextans is at (-36, -53, 58) and Pal 5 is at (7, 0, 16).

Because of the small declination coverage currently available, the
full extent of these large halo structures is not known.  However, an
extent of 30 degrees in RA at 46 kpc, with a `depth' of at up to ten
kpc (from the spread in magnitudes in Figure \ref{fg13}), suggests a
structure of extent at least 20 kpc by at least 2 kpc by up to 10 kpc.
A significant fraction of the assumed width from the magnitude spread
could be due to the intrinsic spread in the absolute magnitudes of BHB
stars, and the actual structures could in fact be extremely long and
narrow ``streams''.  One can estimate the implied total mass for the
large structures in the northern and southern samples from the number of
BHB stars observed.  From the data
associated with Figure \ref{fg13}, one estimates that there are 200
BHB stars (above background) in the northern structure.  To arrive at this
estimate, one uses the counts per square degree of BS and BHB stars in the
left hand panel in Figure \ref{fg13} to estimate the background; the counts
above background are assumed to be associated with the structures.  Excess
counts at $g^* \sim 19$ in the northern sample and at $g^* \sim 18$ in the
southern sample were assumed to be BHB stars.  Likewise, excess counts
at $g^* \sim 21$ in the northern sample and at $g^* \sim 20$ in the
southern sample were assumed to be BSs.

In Pal 5, there are five BHB stars in the sample.  From comparisons of
the total magnitude of Pal 5 to other globular clusters as tabulated
in \citet{dm93}, we estimate the mass of Pal 5 to be $1.5\times
10^5M_\odot$.  Scaling up from a globular cluster mass, a lower limit
on the inferred mass in the part of the northern structure detected is
then $6 \times 10^6M_\odot$.  This is roughly consistent with
estimates of the mass from extrapolations from the number of RR Lyrae
stars.  Ivezic et al. (2000) detect nearly 80 RR Lyrae variables at
$\sim 50$\% efficiency within the northern clump, as compared to five
in Pal 5, giving a ratio of 160 to five as compared with the above
ratio of 200 to five.  An estimated 60 BHB stars in the southern
structure implies a lower limit for the mass of $2 \times
10^6M_\odot$.  There is of course no way to tell what fraction of the
structure is detected; more data from the SDSS survey will eventually
be able to map out the full extent of such structures.

It is interesting to calculate the ratio of BS:BHB stars in the large
structures.  In the globular cluster Pal 5, in the color range $-0.3 <
g^* - r^* < 0$, there are 7 BS stars and 5 BHBs for a ratio of 1.4:1.
In the northern clump, an estimate of 450 BSs is made by assuming a
background similar to that in the upper left hand panel of
\ref{fg13}, giving a ratio of 2.2:1 when combined with the 200 BHBs
seen.  The BSs are near the limit of detection in the north, and the
sample is incomplete, so the ratio may be considerably larger.  For
the southern structure, about 380 BSs are estimated, compared with
about 60 BHBs, for a ratio of 6:1, with considerable uncertainty.  For
A-colored stars in the field not associated with the clumps, a ratio
of about 2:1 is estimated by the same technique.  \cite{pbs94} explore
the populations of blue stars located at $d < 2$ kpc and find a
relatively large specific ratio of blue metal poor to field horizontal
branch objects (which are similar to the BS and BHB categories
characterized here, though based on somewhat different selection
criteria).  This apparent variation in the BS:BHB ratio between globular
clusters, in the field and in the northern and southern
structures, while still somewhat uncertain, appears to be a
significant effect, indicating different populations of stars
populate different regions of the halo.

The large angular extent and relatively low surface brightness of
these structures leads one to believe that these may be tidally
disrupted streamers in the halo.  The fact that two such structures
were detected in 370 square degrees of data ($\sim 1$\% of the sky)
leads one to ask how common these structures might be in the Galactic
halo.  If the distribution of these objects is uniform across the sky,
there may be as many as 200 of these to $20^{th}$ magnitude in $g^*$.
Since the size of the structures parallel to the line of sight could
be as large as 10 kiloparsecs, the area effectively surveyed may be up
to five times larger than the survey area (since the width of the
stripe is approximately 2 kpc at $19^{th}$ magnitude).  In this case
the total expected number of such structures in the halo would be 40,
or fewer if the length of each streamer is much greater than 20 kpc.
The existence of such structures lends strength to the argument that
some or all of the halo of our Galaxy has been formed through the
accretion, or at least the disruption, of smaller stellar systems
\citep{m94}.  Numerous studies have found evidence for coherent
structures in the halo using kinematic information.  \cite{mmh94}
found a clump of 6--9 halo stars about 4--5 kpc above the North
Galactic Pole with very similar radial velocities and proper motions
and identified a moving group.  \cite{hw99} simulate satellite galaxy
disruption and phase-space diffusion in the halo and suggest that the
halo could consist of 300--500 streams with components that pass close
to the sun, a number consistent with that estimated
above. Earlier, \citet{db89} noted a group of 7 field horizontal
branch stars with similar radial velocities all within 1 square degree
on the sky and furthermore noted a positive signal in the 2-point
star-star correlation function of A-colored objects on scales of 10$'$.

A cross-check against 21 cm emitting High Velocity Clouds \citep{wv97} 
does not show any obvious positional matches with these
large structures in the north and south, but if even small amounts of
gas is associated with the structures, a scenario such as that
described in \citet{yetal86} could hold, where the halos of
galaxies are full of small complex, quasar absorption line producing
structures.  If the structures are as numerous as 100 in the halo, then they
could contain a few$\times 10^8 M_\odot$,
a significant fraction of the luminous halo.
This is corroborated by Figure 13, from which one estimates that one in
four of the BHB stars in the combined samples are in one of the two large
clumps (using the color separation of stellar types for field stars, and
magnitude separation of the stellar types in the clumps).

To review: For the structure in the north, one sees BHB and BS
candidate stars and for the structure in the south, which is closer,
one sees these related populations of stars: BHB candidates, BS
candidates, and F stars in numerical proportions which are not
completely out of line with expectations from globular clusters or
that of known galactic populations.  The rough mass estimates for the
structures are somewhat above that for a massive globular cluster, and
only a factor of a few smaller than that of known dwarf companions to
the Milky Way.  Thus galaxies like Ursa Minor or the Draco dwarf
spheroidal at about 70 kpc from the Galactic center, which currently
have central stellar densities much higher than those inferred for the
extended structures in the north and the south, could provide
progenitors of the low density structures seen in Figure \ref{fg3}.
It should be noted that the mass estimates from BHB stars is highly
uncertain, and assumes similar horizontal branch structure for all
populations.  If BS counts were used, the mass estimates would be
several times higher.

\section {FITS TO THE GALACTIC HALO MODEL\label{paramfits}}

The discovery of extensive structures in the outer Galactic halo is
extremely interesting, and the number of A-colored stars associated
with structures is of the same order as that in a supposed smooth
background distribution of halo tracers.  This renders attempts to fit
a smooth spheroid model with a single values of $\alpha$ and $c/a$
problematic.  The approximate separation of BHB and BS stars in
\S\ref{split}, however, can be used to try and isolate only BHB
tracers, and if one restricts that sample further to avoid the obvious
clumps, one may still attempt a significant fit to a smooth spheroid.
Figures \ref{fg20} shows a maximum likelihood fit to flattened
spheroidal models of BHB candidate stars in the North and South.  Fits
were restricted to those BHB color candidates with implied $\rm R <
40\> kpc$, to regions which avoided the clump at $g^*(BHB)= 18 $ in
the south, and to regions which avoided saturated detections of stars
(which occurs in the SDSS at about $g^* \sim 15,\> d \sim 7\>\rm
kpc$).  Constraining the fit to match the numbers of objects seen in
the North and South simultaneously helps greatly in restricting
parameter space.  The model fit uses $\rho \sim R^{\alpha}$,
where $R^2 = (X^2/a^2 + Y^2/b^2 + Z^2/c^2)$, and $X,Y,Z$ are
Galactic rectangular coordinates.  Setting $a=b=1$, $\alpha$ and $c$
are allowed to vary.

The maximum likelihood estimate of the halo structure parameters for
$\rm 7 < d<40 \>kpc$ is: $\alpha = -3.2\pm 0.3,\> c/a = 0.65 \pm 0.2$,
in agreement with \citet{gfb98} and many other authors.  Models with
$c/a = 1.0$ or $\alpha < -2.8$ yield significantly worse fits to the
sample.  After the numerous cuts placed on the input candidate sample,
the number of stars fit is 527 stars.  The fit can be extrapolated to
provide a local density for BHB stars at a distance of 8 kpc from the
Galactic center.  This normalization yields $\rm \rho_{BHB}(8\>
kpc) = 8\> kpc^{-3}$, in reasonable agreement with earlier studies
\citep{ksk94}.


More data, and cleaner spectroscopic separation of BHB
from BS stars will of course help improve the confidence in these
numbers to $r > 40 $ kpc, though if the presence of enormous
structures in the outer halo at larger radii is pervasive, fitting a
single set of parameters to the shape of the outer halo may not
provide much insight into its overall mass distribution.

\section {CONCLUSIONS\label{conclusions}}

\begin{itemize}

\item[1.] A sample of 4208 A-colored stars from the SDSS is
	uniform to $g^* = 19.5$ and extends to $g^* \sim 22$.  This
	sample is used to probe halo structure at distances of up to
	$60 $ kpc from the Galactic center.

\item[2.] A plot of all A-colored stars in magnitude-angle space shows
	significant clumpiness, including a large structure in the
	North at ($\alpha$, $\delta$) = (220$^\circ$, $0^\circ$)
	extending over 30 degrees in length and at least 2.5 degrees
	in width.  The remarkable detection of a clump of 80 RR Lyraes
	in this same area of sky by \citet{ietal00} demonstrates the
	coherence and existence of this structure.  In the South, a
	second extended structure of over 20 degrees in length is
	apparent.  Numerous smaller structures, including the known
	dwarf irregular Sextans (at 75 kpc from the Galactic center) and
	the Globular cluster Pal 5 are also clearly visible.

\item[3.] The example of Pal 5, which has a BHB and a BS sequence two
	magnitudes below its BHB, suggests an explanation for the
	parallel magnitude-angle-space arcs seen in the larger blue
	star halo samples.  Photometric separation of A star surface
	gravities supports this explanation.  A majority of the high
	surface gravity A-colored stars in the halo may be BSs.

\item[4.] Based on followup spectroscopy of eight objects, there are at
	least two sub-populations of A-colored stars identified: blue
	horizontal branch stars with low surface gravities,
	and blue straggler stars with high surface gravities
	and absolute magnitudes one to three magnitudes fainter than
	the horizontal branch.  

\item[5.] The assumption of $M_{g^*} \sim 0.7$ for the horizontal branch
	yields approximate distances to the large
	clumps in the North and South of 46 kpc and 33 kpc from the
	Galactic center at positions (X,Y,Z) = (21, -5, 38) and
	(-22, 6, -24) kpc, respectively.  The inferred masses of the
	detected structures is at least $6 \times 10^6M_\odot$ in the
	north and at least $2 \times 10^6M_\odot$ in the south, though
	it is not known what fraction of the total extent of these
	elongated structures has so far been detected.  From estimates
	of the mass in these structures and the fraction of sample
	BHB stars they contain, one concludes that structures such as
	these could contain a significant fraction of the mass of the
	luminous halo.

\item[6.] Fitting the `field' BHB candidate stars, (those which avoid
	the structures), while requiring a consistent normalization
	between northern and southern Galactic samples, yields a
	maximum likelihood fit to a spheroidal halo model of $\alpha =
	-3.2\pm 0.3$ and flattening of $c/a = 0.65\pm 0.2$ for $7 < R
	< 40$ kpc, consistent with previous work.  Since clumps of
	A-colored objects dominate the numbers of halo BHBs and BSs at
	distances $> 30 $ kpc, however, simple parameter fits to single
	spheroid models are not representative of the whole picture.
	The existence of enormous structures in the outer halo must be
	addressed.

\end{itemize}

\acknowledgments

We acknowledge useful discussions with R. Kron, B. Gibson and many
important suggestions and comments from the referee, T. Beers.
We thank R. McMillan and K. Gloria for excellent assistance in
operating the APO 3.5m telescope.

This work was supported by Fermi National Accelerator Laboratory,
under U.S. Government Contract No. DE-AC02-76CH03000.

SALM acknowledges partial support from the Department of Energy at the
Lawrence Livermore National Laboratory under contract W-7405-ENG-48 and
from the NSF (grant AST-98-02791).

The Sloan Digital Sky Survey (SDSS) is a joint project of The
University of Chicago, Fermilab, the Institute for Advanced Study, the
Japan Participation Group, The Johns Hopkins University, the
Max-Planck-Institute for Astronomy, Princeton University, the United
States Naval Observatory, and the University of Washington. Apache
Point Observatory, site of the SDSS, is operated by the Astrophysical
Research Consortium. Funding for the project has been provided by the
Alfred P. Sloan Foundation, the SDSS member institutions, the National
Aeronautics and Space Administration, the National Science Foundation,
the U.S. Department of Energy, and Monbusho. The SDSS Web site is 
http://www.sdss.org/.

\clearpage

\figcaption {$u^*-g^*$ vs. $g^*-r^*$ color-color diagram for SDSS
commissioning data on the celestial equator.  The plot shows the
colors of $\sim 30,000$ stars from 13.3 square degrees of
commissioning data in the range $\rm 165^\circ <\alpha< 225^\circ$, $\rm
-0.23^\circ < \delta  < 0.0^\circ$.  This represents 4 hours of
observations from one of six columns of CCDs in the imaging camera.
Each column of CCDs images a strip of sky 0.23 degrees wide in five
passbands.  Objects in different magnitude ranges are shaded
differently.  The A-colored star selection box is identified, as are
boxes for quasar candidates, F, and G type colored stars.
\label{fg1}}

\figcaption { The upper line indicates fraction of stars matched
between runs 745 and 756 without any restrictions on color.  This
indicates a detection limit of about $g^* \sim 22.5$ (50\%).  The
lower line shows the fraction of stars in the A star color-color box
($0.8 < u^*-g^* < 1.5, -0.3 < g^*-r^* < 0.0$) that are matched as a
function of $g^*$ magnitude.  This indicates a uniform completeness
limit of about $g^* \sim 20$ in the northern sample.  The low matched
fraction (80\%) at bright magnitudes is due to color errors at the red
end of the box ejecting one or another detection of a star from the
matched pair out of the color box.
\label{fg2}}

\figcaption {A polar wedge plot of RA vs. $g^*$ magnitude for
A-colored stars.  Notice the parallel arcs at $ 195^\circ <\alpha<
230^\circ$ and the large structure at $20^\circ <\alpha< 40^\circ$, as
well as the Sextans dwarf irregular galaxy at $\alpha=153^\circ$,
$g^*=20.5$ and the Pal 5 BHB at $\alpha=229^\circ$, $g^* = 17.3$.  The
Pal 5 blue straggler sequence is visible as a thin line at
$\alpha=229^\circ$, $18.5 < g^* < 20.5$.  The intersection between
the plane of the celestial equator and the plane of the Galaxy is
indicated by a dark line.  Galactic and celestial coordinates are
indicated around the edge of the figure. $g^*$ magnitude increases
outward from the center.  Pal 5 is notable in the enlarged in the
small diagram to the right, rotated 90$^\circ$ counterclockwise from
the main figure.
\label{fg3}}

\figcaption {Wedge plot of RA vs. $g^*$ magnitude for color-selected
quasar candidates (objects in the small quasar box of Figure
\ref{fg1} with $-0.05 < u^*-g^* < 0.15, -0.1 < g^*-r^* < 0.0$).  This
population of almost exclusively extragalactic objects is very
uniformly distributed in RA to at least $g^* = 22$ (where there is
contamination from F stars with large color errors). This demonstrates
the uniformity of the SDSS data sample.  The Northern sample reaches
about 0.5 mag deeper than the Southern sample due to increased $u^*$
errors for the somewhat lower quality Southern sample data leading to
a brighter $u^*_{\rm err} < 0.30$ cutoff in the South.
\label{fg4}}

\figcaption {Number counts of quasars in the North and South samples.
Densities of extragalactic candidate objects in the large quasar
candidate color-color box of Figure \ref{fg1} ($-0.05 < u^*-g^* <
0.35, -0.1 < g^*-r^* < 0.3$) in the North and South are overplotted.
The close agreement between the two curves demonstrates the
consistency of the SDSS photometric calibration for this isotropically
distributed sample of objects.  The magnitude limit in the North is
about 0.5 m fainter than that in the Southern sample.  The Northern
sample reaches about 0.5 mag deeper than the Southern sample due to
increased $u^*$ errors for the somewhat lower quality Southern sample
data leading to a brighter $u^*_{\rm err} < 0.30$ cutoff in the South.
\label{fg5}}

\figcaption {Wedge plot of RA vs. $g^*$ magnitude for A-colored stars
selected in independent SDSS runs. Similar to Figure \ref{fg3}, the
reality of clumped structures is independently verified. This data set
doesn't cross the longitude of Sextans, and since it is only data for
one-half of a filled stripe, it only has half the stellar density of
the Figure \ref{fg3}.  However, the coverage makes the end of the
northern structure is more apparent than in Figure \ref{fg3}.
\label{fg6}}

\figcaption {Color-magnitude diagram around the globular cluster Pal
5.  The blue A-colored object box limits in $g^*-r^*$ are indicated by
two vertical lines.  Note the 5 BHB stars and 7 candidate blue
straggler stars, separated by about 2 magnitudes in $g^*$.
The circled objects indicate the BHB and BS objects. Objects
with squared circles were followed-up spectroscopically.
\label{fg7}}

\figcaption {Spectra of a Pal 5 BHB and Pal 5 BS star. The upper
panel shows the Balmer lines for a horizontal branch star in the
cluster Pal 5.  The width of the $H\delta$ and $H\gamma$ lines are
clearly better matched to the reference BHB star (a $V = 12$ field
BHB) than to the reference field A0V star.  The horizontal line marks
80\% of the continuum of the stars, facilitating the quantitative
width measurement.  The lower panel shows a Pal 5 blue straggler
spectrum, which clearly fits the high gravity reference A0V star.
\label{fg8}}

\figcaption {Six ARC 3.5m spectra of SDSS A-colored stars.  The star in the
lower left is best fit by a high gravity model while the other 5 are
narrower and consistent with BHB gravities. The star in the upper left
has a narrower line width, and the fluctuating continuum suggests
a metal-line enhanced Am star classification.
\label{fg9}}

\figcaption {$u^*-g^*$ vs. $g^*-r^*$ color-color plot for stellar
models and stars with measured surface gravities.  Models from Lenz et
al (1998), valid for stars with $6500 < T_{\rm eff} < 10000 K$, are
plotted in small filled circles (low gravity BHB models) and small
open circles (high gravity BS/A models).  The range of colors for each
model represents a range of metalicities.  The eight stars for which
spectra were obtained plus other Pal 5 BHBs and BSs are plotted on the
diagram as triangles and squares.  The heavy black line then
represents an empirical separation curve between stars of high and low
gravity, and is used to separate the SDSS A-colored star sample into
BS candidates and BHB candidates based on colors alone.  A few (low-gravity)
variable RR Lyraes from \citet{ietal00}, put on the color system of
this paper, are also indicated.
\label{fg10}}

\figcaption {Wedge plot of RA vs. $g^*$ magnitude for stars with
colors of candidate BHBs.  The well defined horizontal
branches of the arcs at $g^* = 19 $ in the North and $g^* = 18$ in
the south, as well as Pal 5's BHB, clearly stand out in this selection of
photometrically selected BHB candidates.
\label{fg11}}

\figcaption {Wedge plot of RA vs. $g^*$ magnitude for stars with colors of
BSs.  The distant (fainter) clumps are defined
in intrinsically fainter BS candidate A-colored stars, photometrically
separated from the BHBs.
\label{fg12}}

\figcaption {Magnitude histograms for candidate BHB and BS stars in
the northern and southern structures.  The northern diagrams include
all blue stars with $\rm RA < 200^\circ$ (left panels), and $\rm RA >
200^\circ$ (middle panels), and the southern diagrams include all
stars with $\rm RA > 0^\circ$.  There are clearly clustered horizontal
branch objects at $g^*= 19$ in the North at $RA > 200^\circ$, and at $g^*
= 18$ in the South.
\label{fg13}}

\figcaption { $g^*-r^*$ color histogram of all objects in the northern
structure (upper panel) with BHB candidate colors and $18.8 < g^* <
19.4$ (thick line) and BS candidate colors and $g^* > 20$ (thin line).
The southern structure histograms (lower panel) are objects in the
clump with $17.7 < g^* < 18.3$ for BHB candidates (thick line) and $19
< g^* < 20.5$ for BS candidates (thin line). Note that most of the BS
candidates are on the red side of the diagram.
\label{fg14}}

\figcaption {Wedge plot of RA vs. $g^*$ for a simulated smooth galaxy
halo of $M_{g^*} = 0.7$ BHBs with $\alpha = -3$ and $c/a = 1$.  The
entire celestial circle is shown (not just for regions where SDSS data
exists).  The maximal density of stars is nearly, but not exactly, at
$b=0$.  Such a density maximum is a poor fit to the clump seen in
Figure \ref{fg3} (with maximum density at $l\sim 0$, and no
equivalent maximum is seen in the south at $l=180$.  This is further
evidence that the structures seen in Figure \ref{fg3} are real halo
substructures.
\label{fg15}}

\figcaption {Wedge plot of RA vs. $g^*$ magnitude for F star
candidates (objects in the box $0.6 < u^*-g^* < 1.0,\> 0.1 < g^*-r^* <
0.3$.  Because of the high relative density of these objects a randomly
selected subsample of 1 in 10 points is actually plotted. The F stars
in the South at $\rm 20^\circ <\alpha< 40^\circ$ at $g^* =21.5$ are
apparently associated with the A-colored star clump.  In the North,
the Galactic halo distribution overwhelms any clumps.
Clump-associated F stars in the North would have magnitudes $g^* \sim
22-23$.
\label{fg16}}

\figcaption {Wedge plot of RA vs. $g^*$ magnitude for G star
candidates ($0.9 < u^*-g^* < 1.3,\> 0.3 < g^*-r^* < 0.4$). Because of
the relative density of these objects, a randomly selected subsample
of 1 in 5 points is actually plotted. The structure of the disk is
becoming manifest as one move to lower Galactic latitudes.  No trace
of the clumps visible in A-colored objects remains for G stars of
similar magnitudes.  Clump-associated G stars would be at $g^* > 24$.
A few F stars may be leaking into the G star box, as indicated by the
presence of excess stars near $g^*= 21.5$ in the south.
\label{fg17}}

\figcaption {Wedge plot of RA vs. inferred Galactocentric radius for
candidate BHB stars. An average absolute magnitude of $M_g = +0.7$ is
assumed to infer distances to the BHB photometrically selected
sample.  Horizontal branch concentrations are visible at $\rm
R=46\>kpc$ in the North and $\rm R=33\> kpc$ in the South.  Sextans is
visible at about 80 kpc.  Other structures, such as the one at
$\alpha=194^\circ$, R=50 kpc remain to be confirmed.  The limiting
magnitude ($g^* \sim 22.5$) allows detection of BHBs out to 150 kpc,
but the sample at the faintest limits is contaminated by intrinsically
fainter BSs, and so the objects apparently out at $R > 75\>$ kpc are not
confirmed to actually be at these great distances.  \label{fg18}}

\figcaption {Wedge plot of RA vs. inferred Galactocentric radius for
candidate BS stars.  An average absolute magnitude of $M_g = +2.7$ is
assumed to infer distance to the BSs.  Loose concentrations
at $\rm R=46\> kpc\>$ in the North and $\rm R=30\> kpc\>$ in the South
are visible. The limiting magnitude cuts off the distribution at about
60--70 kpc.  \label{fg19}}

\figcaption { Binned representations of the data and best fit
solutions for a simultaneous two parameter fit (plus normalization to
the total number of stars) to the stars in the northern and southern
BHB candidate (color selected) subsamples after the stars in the large
structures have been removed.  The panel on the upper left shows the
northern sample star counts vs. distance from the sun and a binned
likelihood fit model.  The upper right panel shows the fit and star
count histograms vs. RA.  The error bars on the fit represent Poisson
statistics for the number of expected objects in the fitted bin.  For
display purposes, the bin size is 2 kpc in d and 5 degrees in RA. The
southern subsample and fits are shown in the lower two panels.  The
maximum likelihood fit yields $\alpha = -3.2 \pm 0.3$ and $c/a =
0.65\pm 0.2$. \label{fg20}}

\begin{table}
\begin{tabular}{lllllrrr}
\multicolumn{8}{c}{Table 1 -- Spectroscopically targeted A-colored stars} \\
\hline
\hline
\multicolumn{1}{c} {ID} &
\multicolumn{1}{c} {$g^*$} &   
\multicolumn{1}{c} {$u^*-g^*$} &   
\multicolumn{1}{c} {$g^*-r^*$} &   
\multicolumn{1}{c} {$H_{\delta}(0.8)$[\AA]} &   
\multicolumn{1}{c} {$l$ (deg)} &   
\multicolumn{1}{c} {$b$ (deg)} &   
\multicolumn{1}{c} {Class} \\
\hline
CS29516$-$0011$^a$&12.24$^b$ &1.30$^b$ & -0.17$^b$&21&71 &-42  & BHB\\
HD13433	 &$\>\>$8.16$^b$&1.10$^b$ &-0.14$^b$ &35 &184 &-68  & A0V\\
WD0148p467$^c$ &12.44$^b$&0.20$^b$ &-0.17$^b$ &60 &134 &-15& DA WD\\
SDSSp J151610.80$-$000745.0&17.20&1.13 & -0.20&25  & 1 & 46  & BHB$^d$\\
SDSSp J151608.00$-$000821.0&18.88&1.12 &-0.13  & 36 &1 &46 &  BS$^d$ \\
\hline
SDSSp J154256.50$-$005609.0 &17.55& 0.99& -0.29&31 & 6 &40 &  BHB\\
SDSSp J162341.10$+$000926.0 &17.52& 1.18& -0.24& 26 &14 &32 &  BHB \\
SDSSp J160326.10$+$005104.0 &17.49& 1.18& -0.12 &  27 & 11& 37  &  Am\\
SDSSp J162520.10$+$000223.0 & 17.49& 0.96& -0.26 &33 & 15&32&  A/BS \\
SDSSp J162025.50$-$004537.0 & 17.82& 1.11&-0.23 & 26 &13&33&  BHB\\
SDSSp J162532.10$-$011215.0 & 17.54& 1.11&-0.20& 26 &13&31&  BHB\\
\hline
\end{tabular}

\begin{itemize}
\item[$^a$]{Ref. Wilhelm et al. 1999}
\item[$^b$]{Magnitudes and colors converted from measured $U, B, \> V$ values}
\item[$^c$]{Ref. Gliese \& Jahreiss 1979}
\item[$^d$]{These objects are in the Globular cluster Pal 5}

\end{itemize}
\end{table}

\clearpage

\setcounter{page}{0}

\plotone{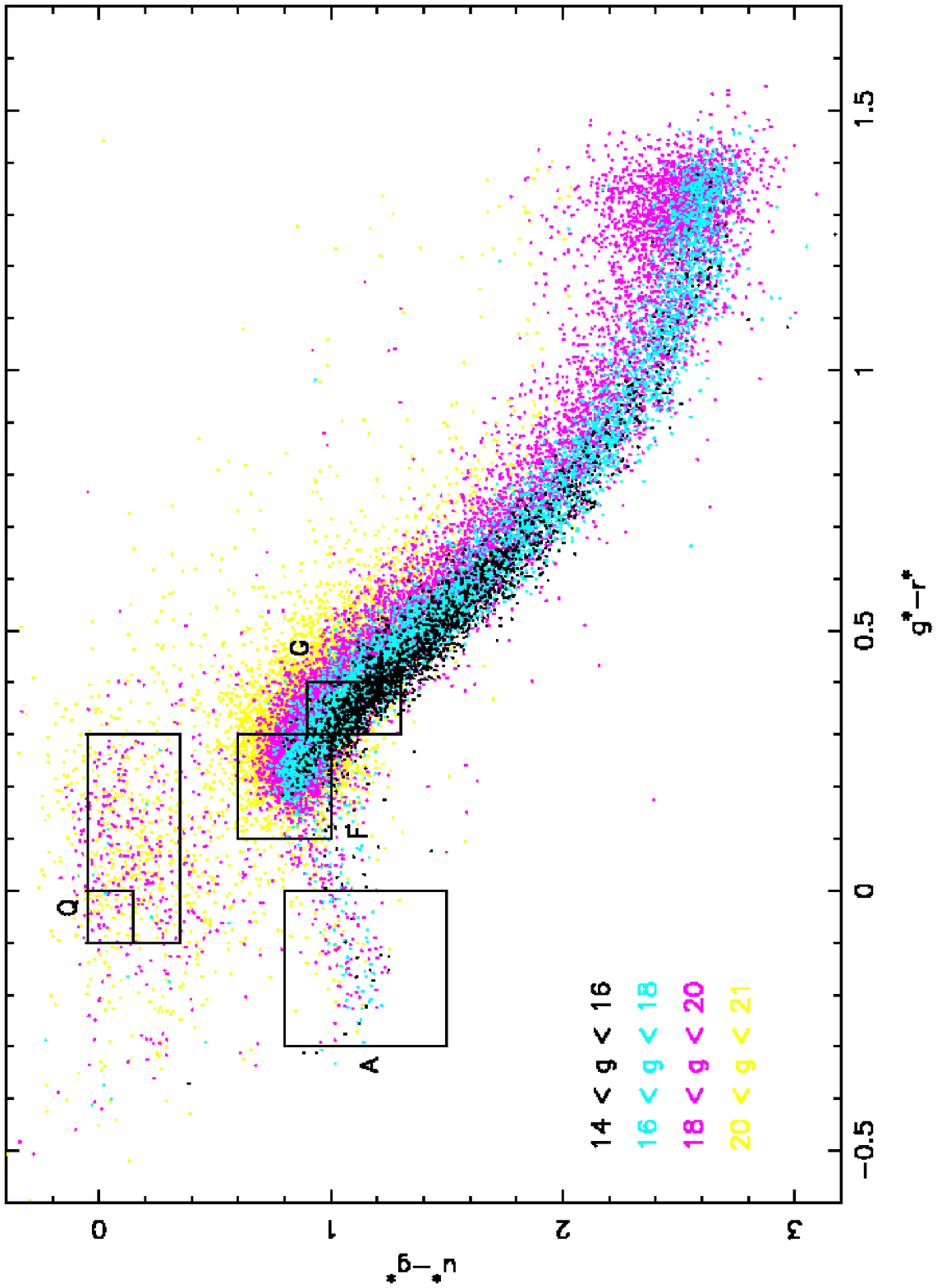}
\plotone{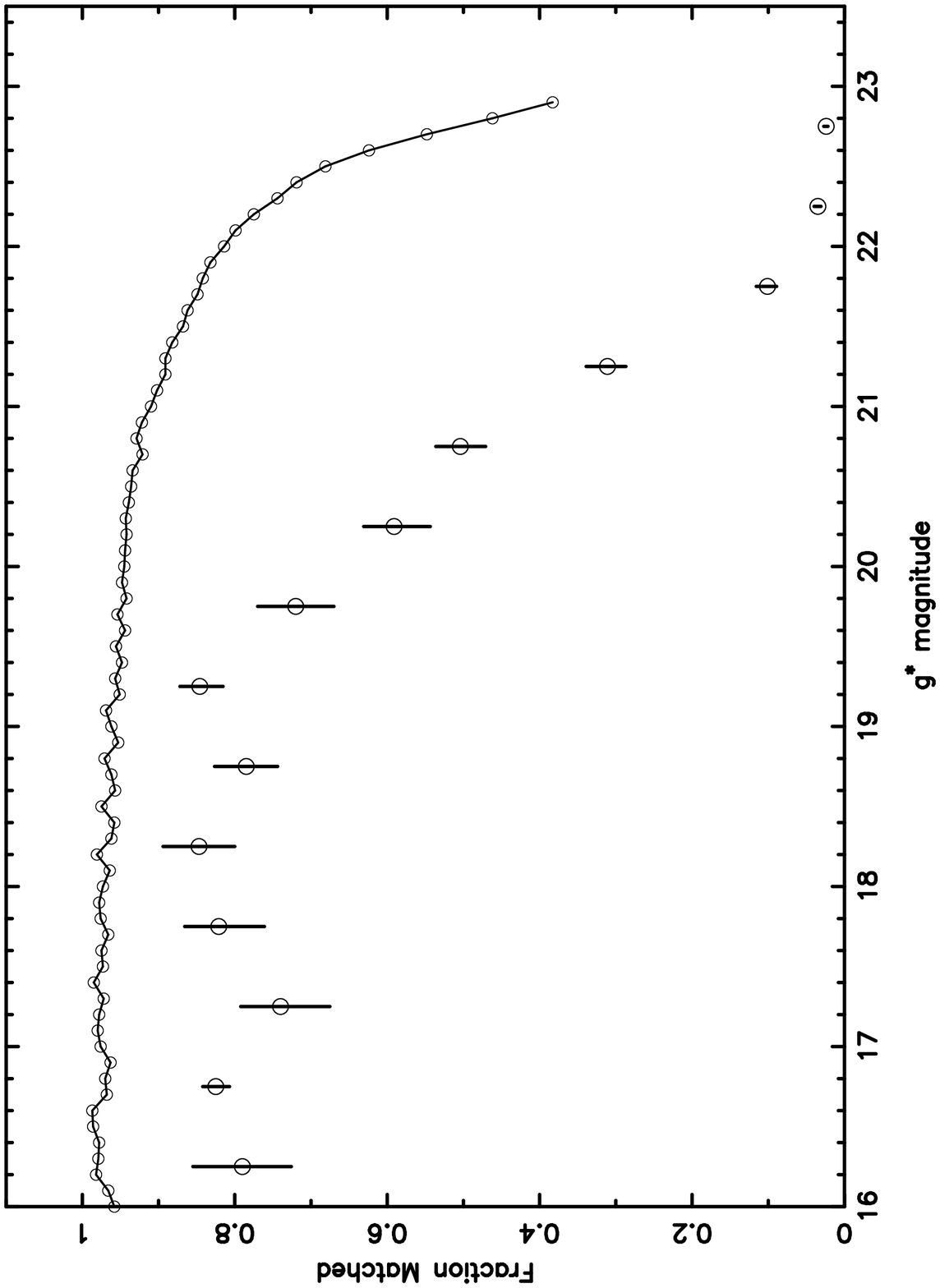}
\plotone{yfig3.ps}
\plotone{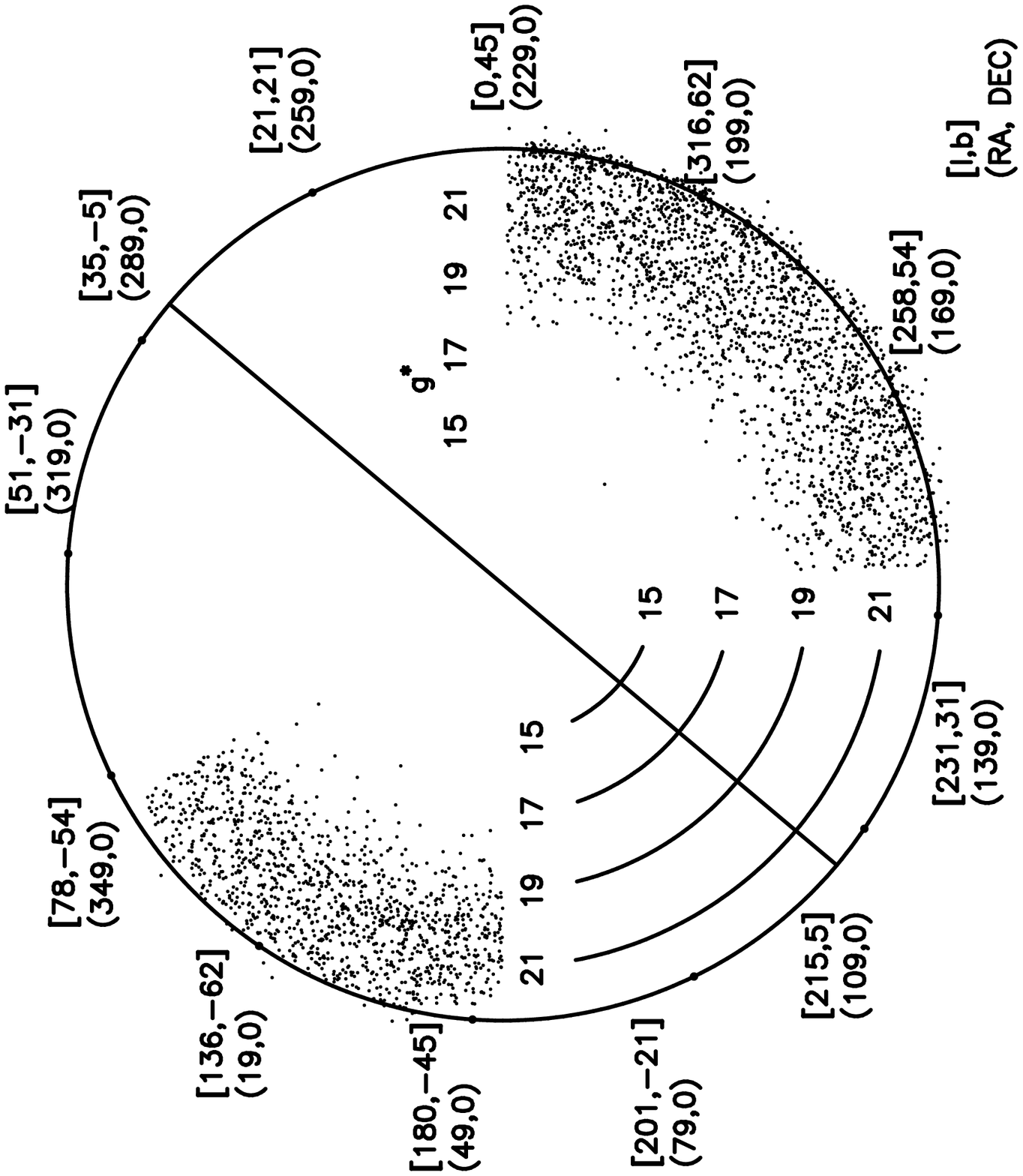}
\plotone{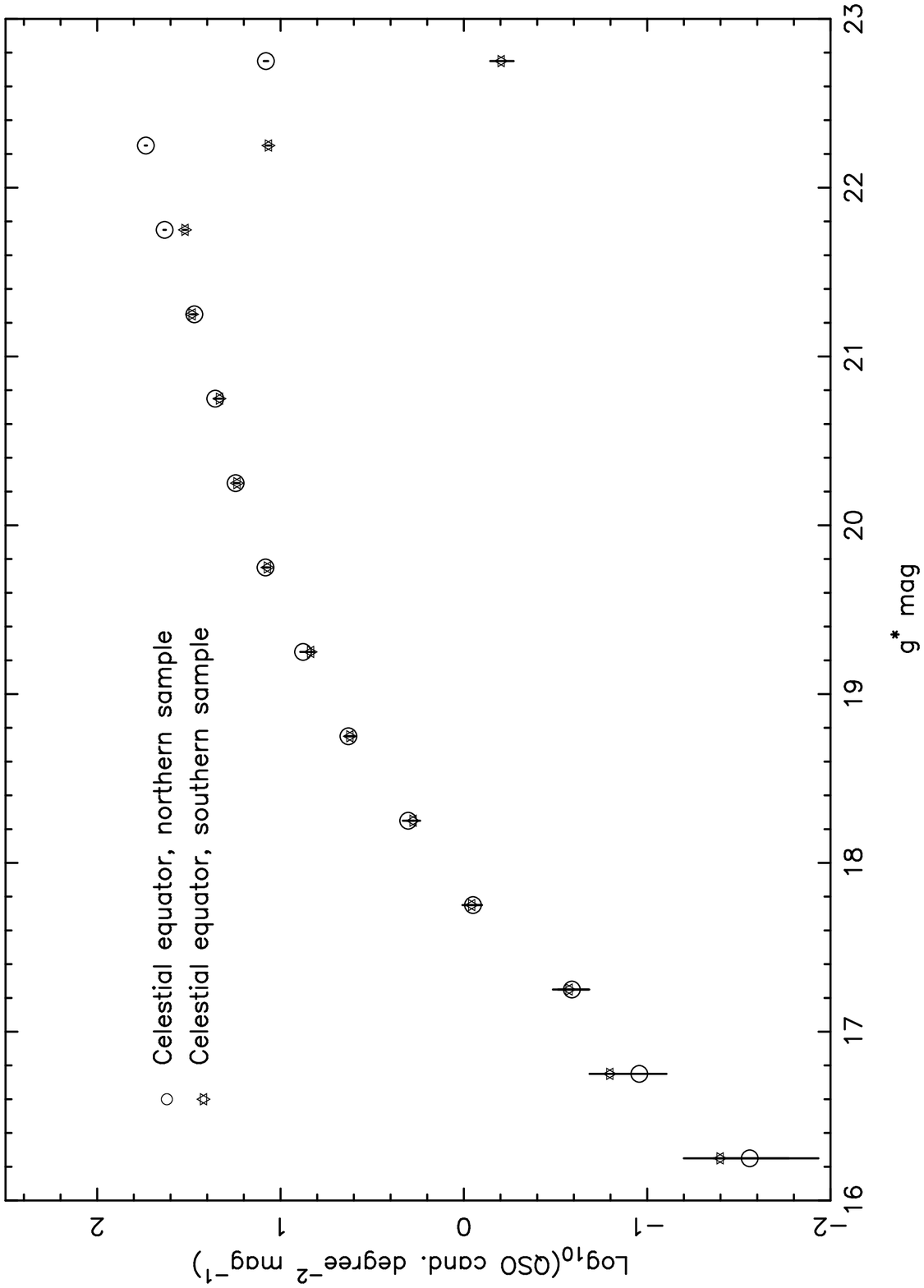}
\plotone{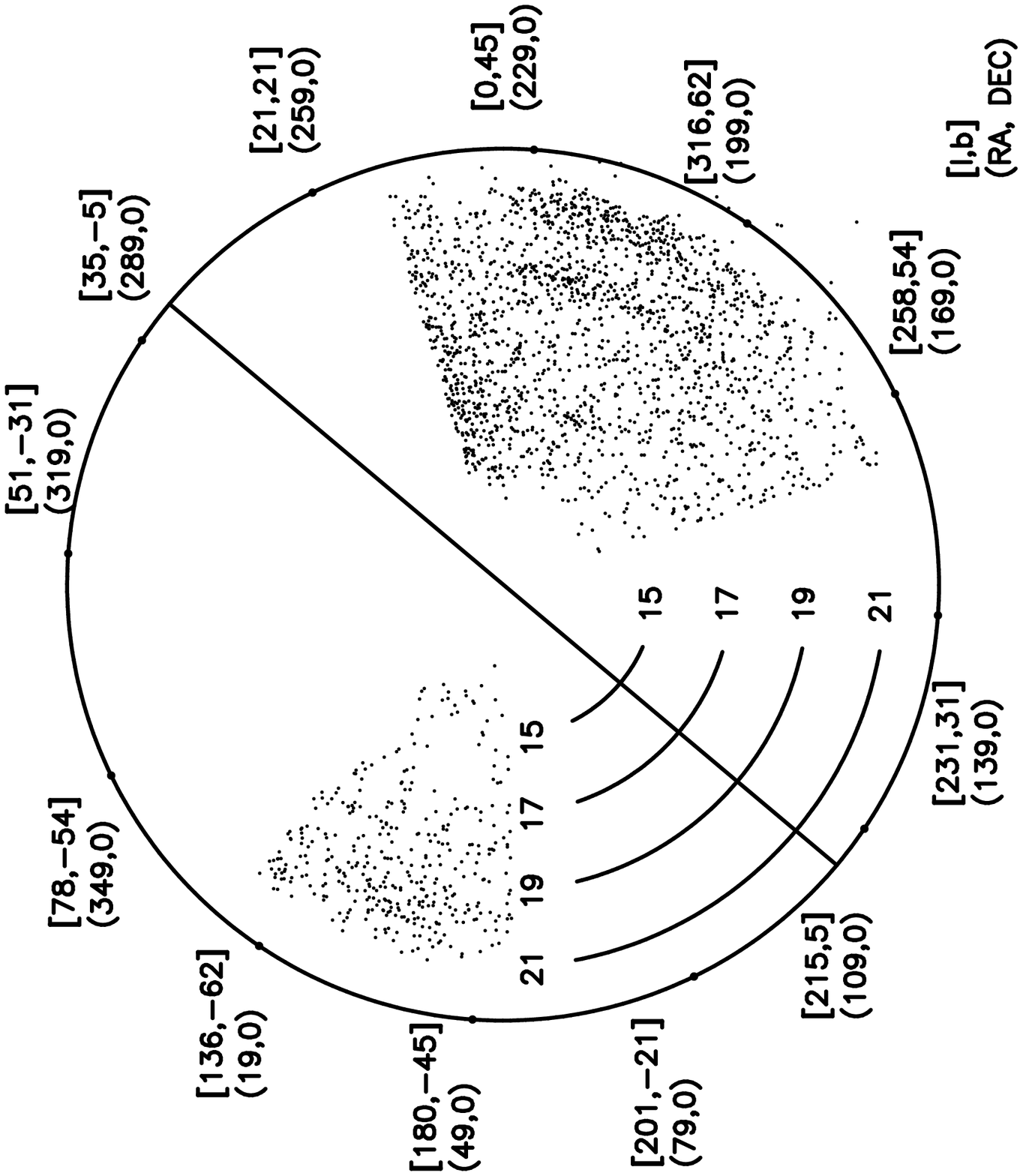}
\plotone{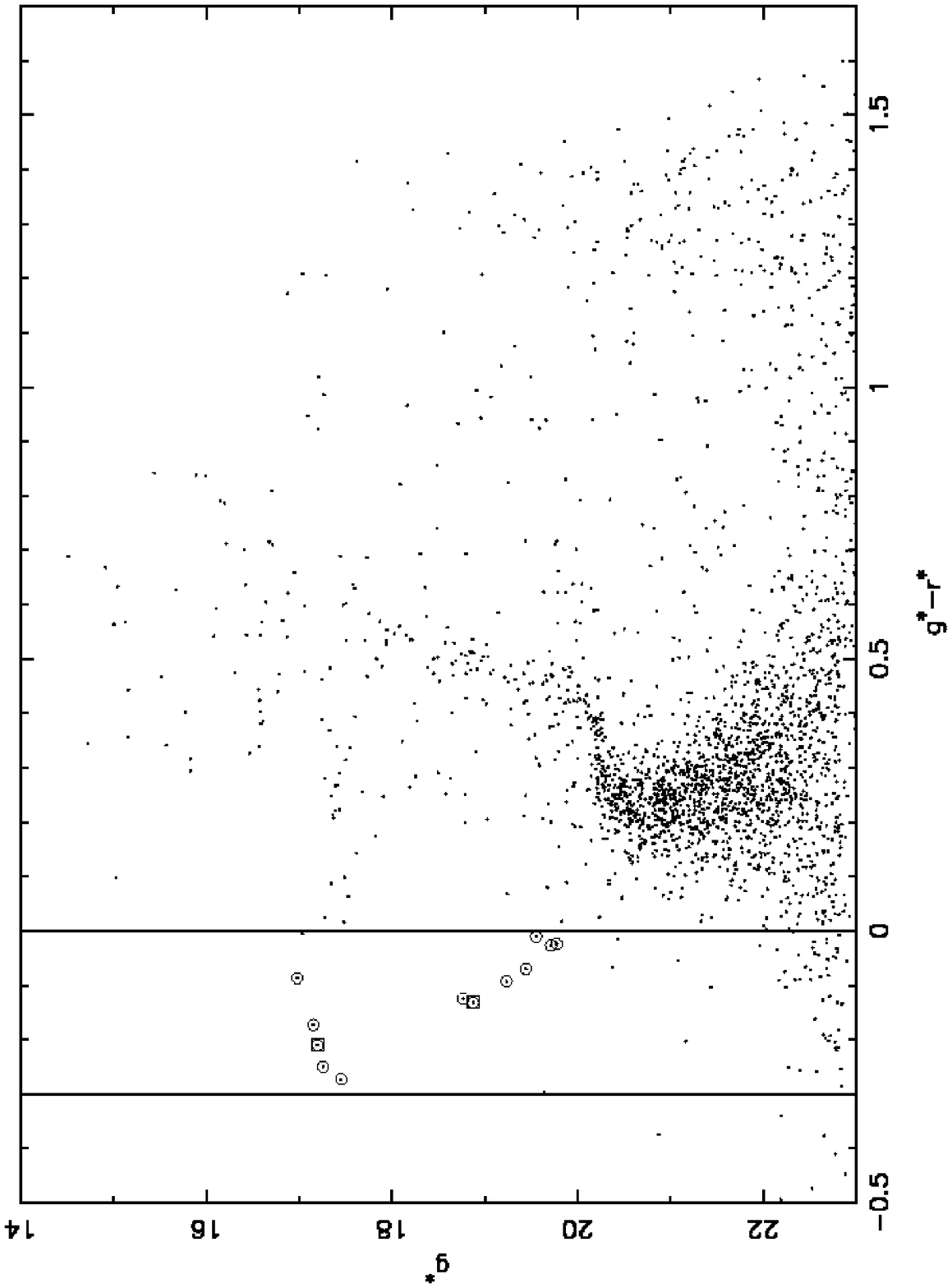}
\plotone{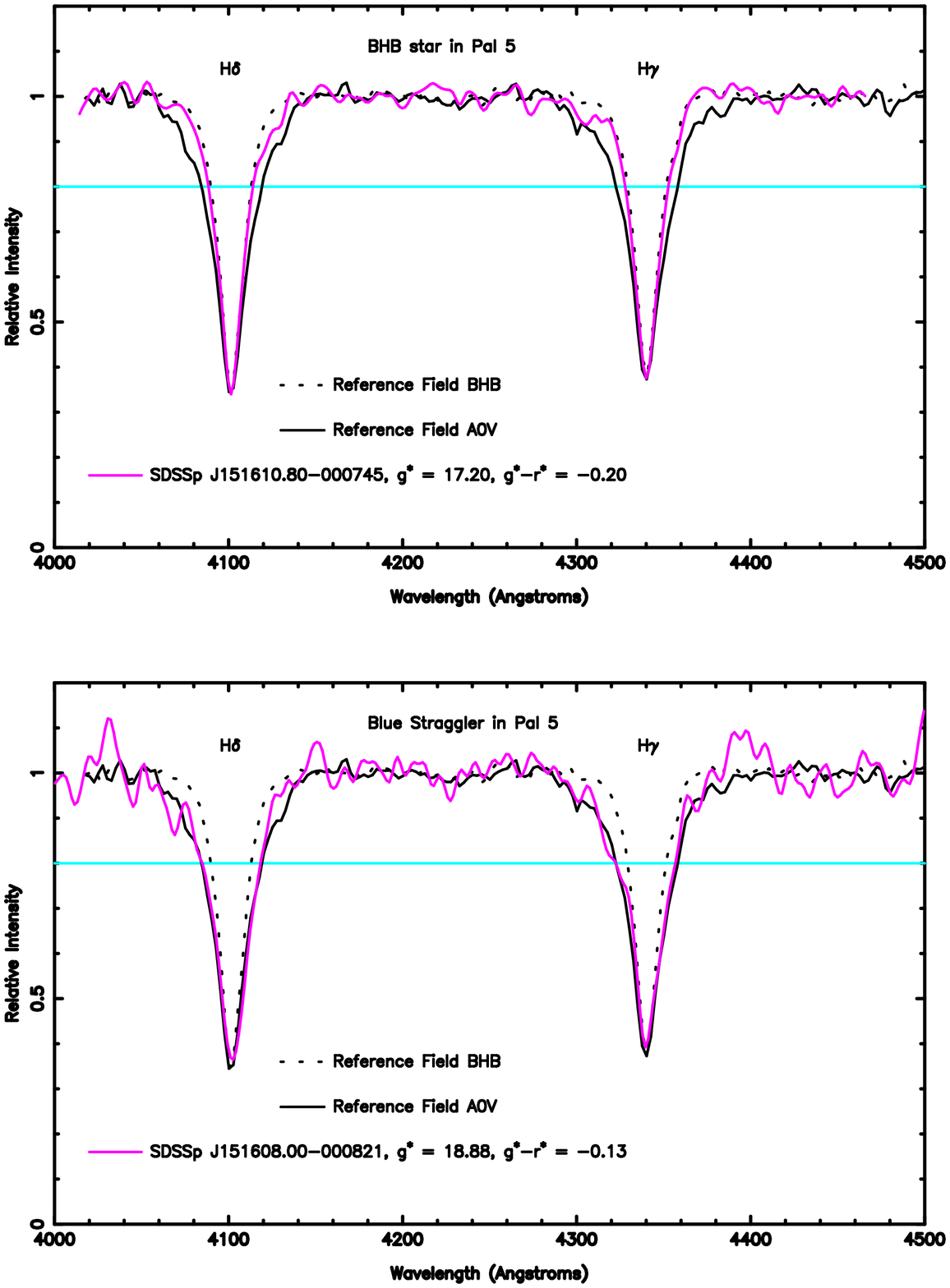}
\plotone{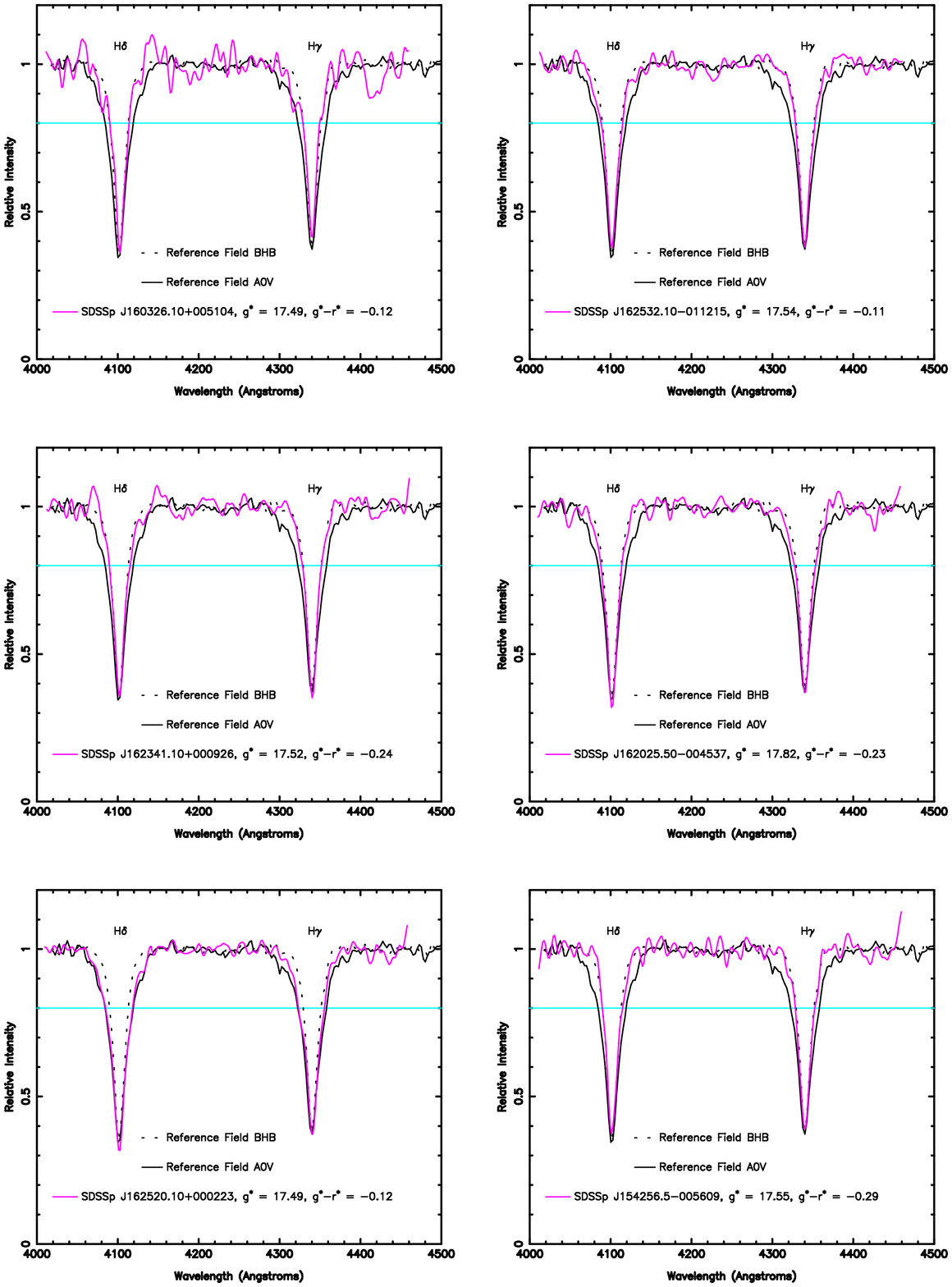}
\plotone{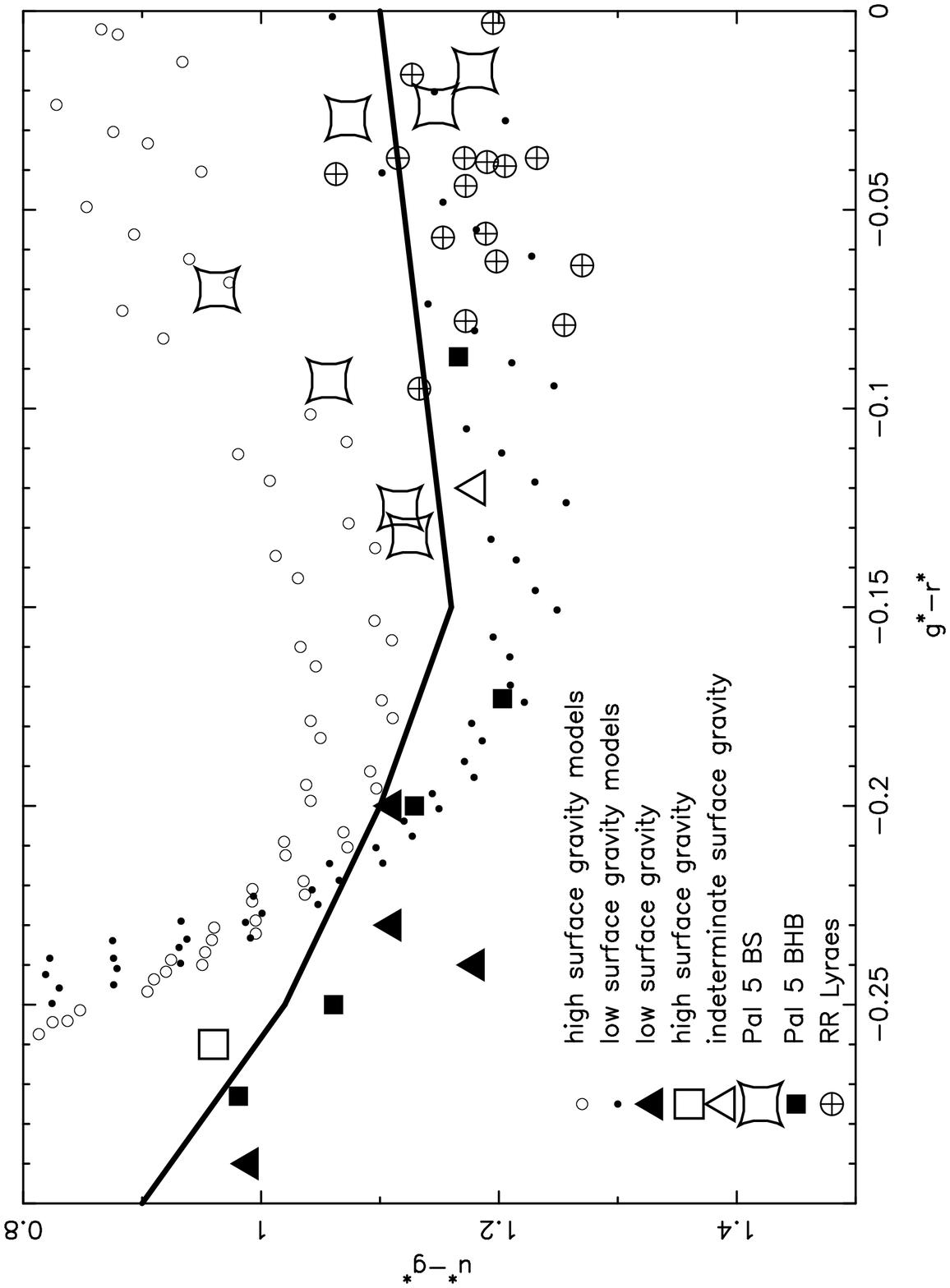}
\plotone{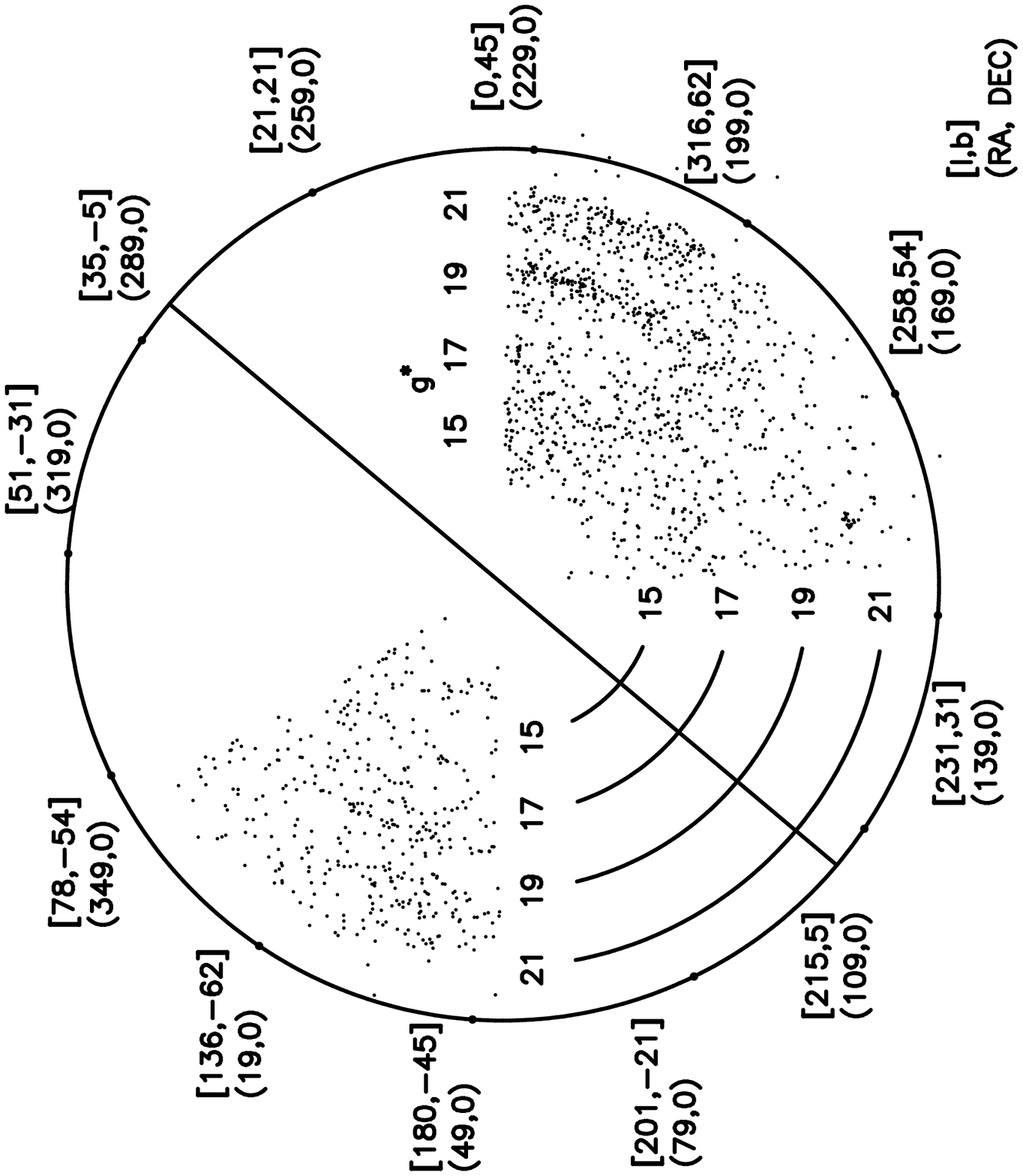}
\plotone{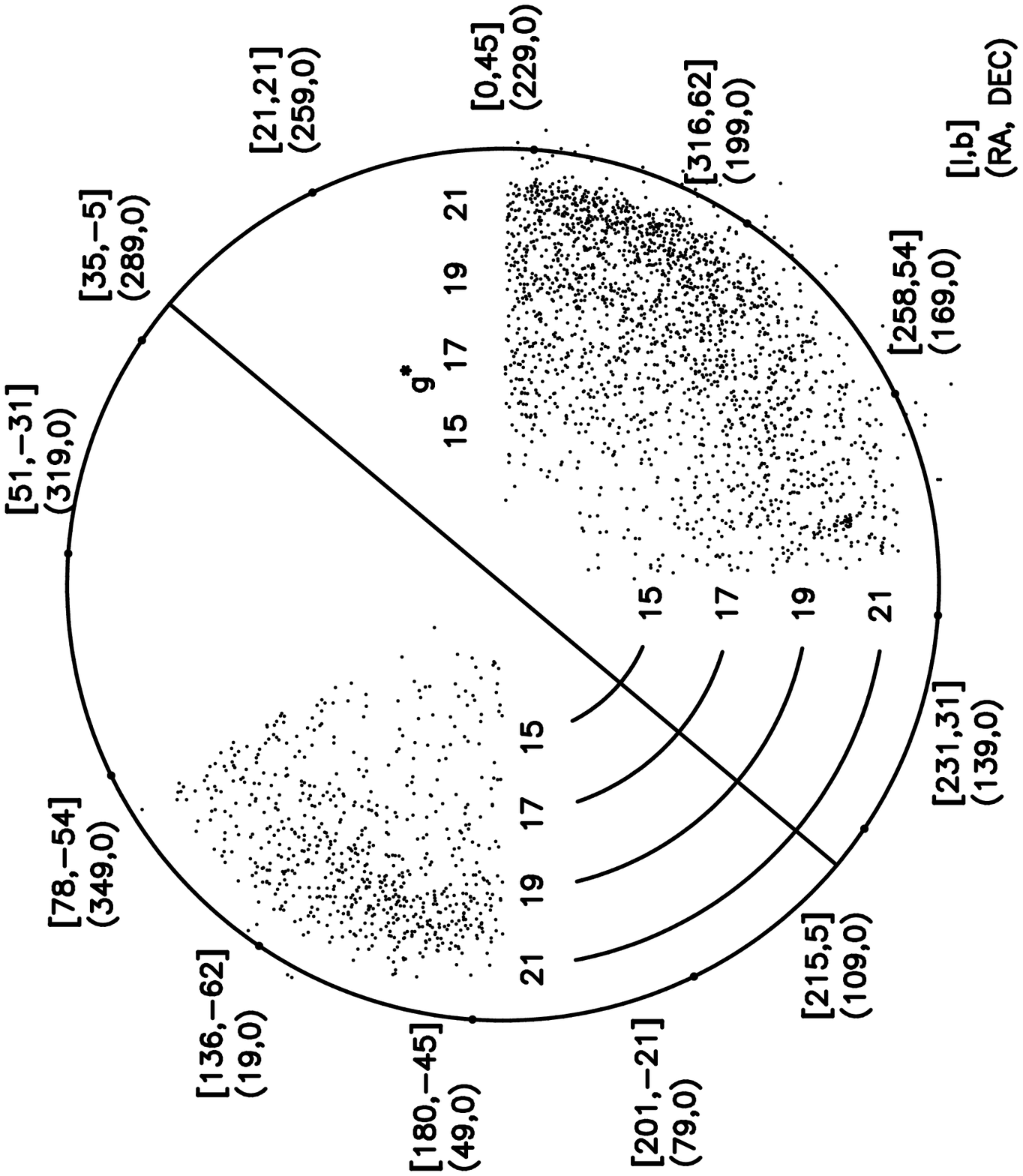}
\plotone{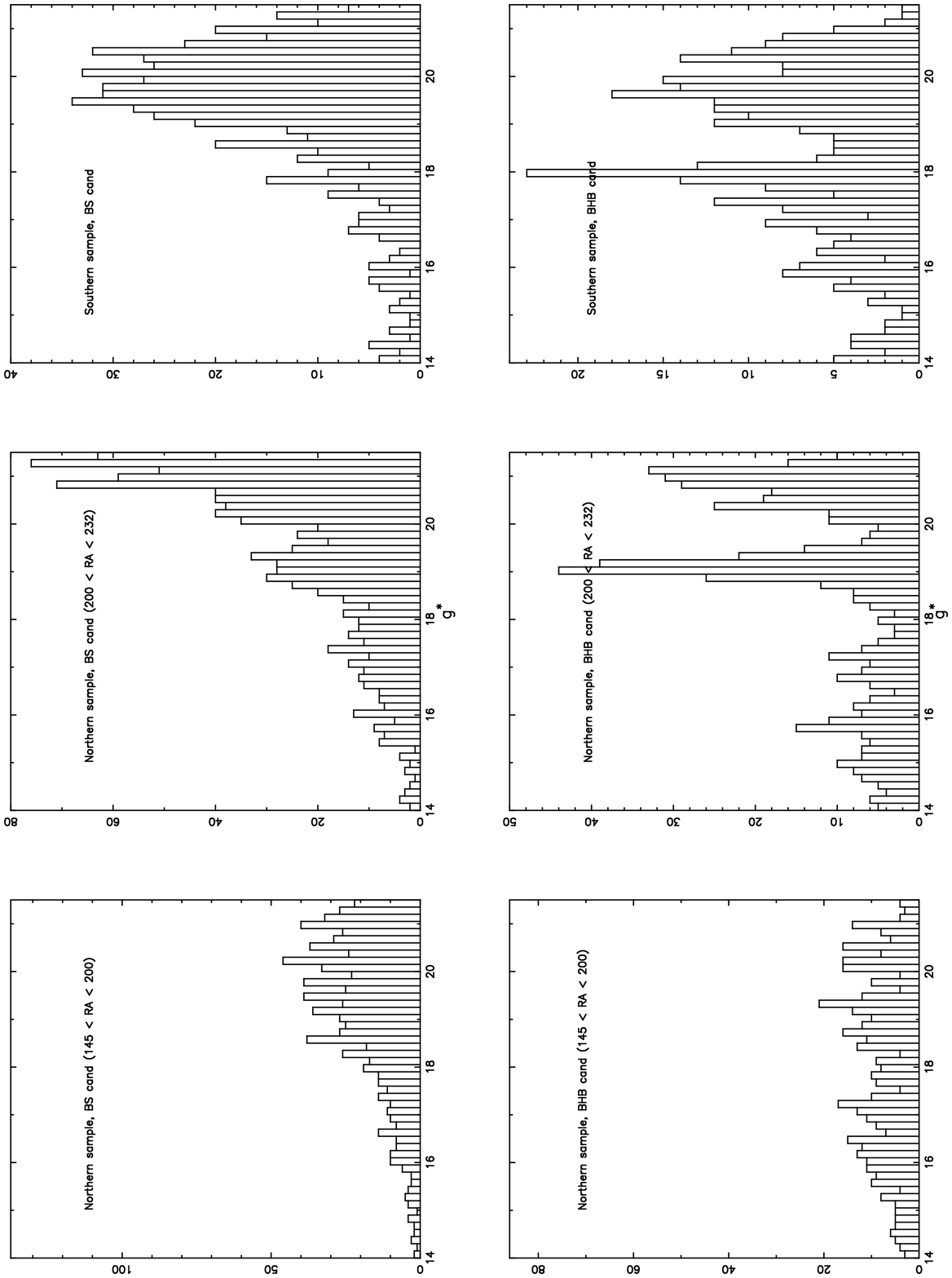}
\plotone{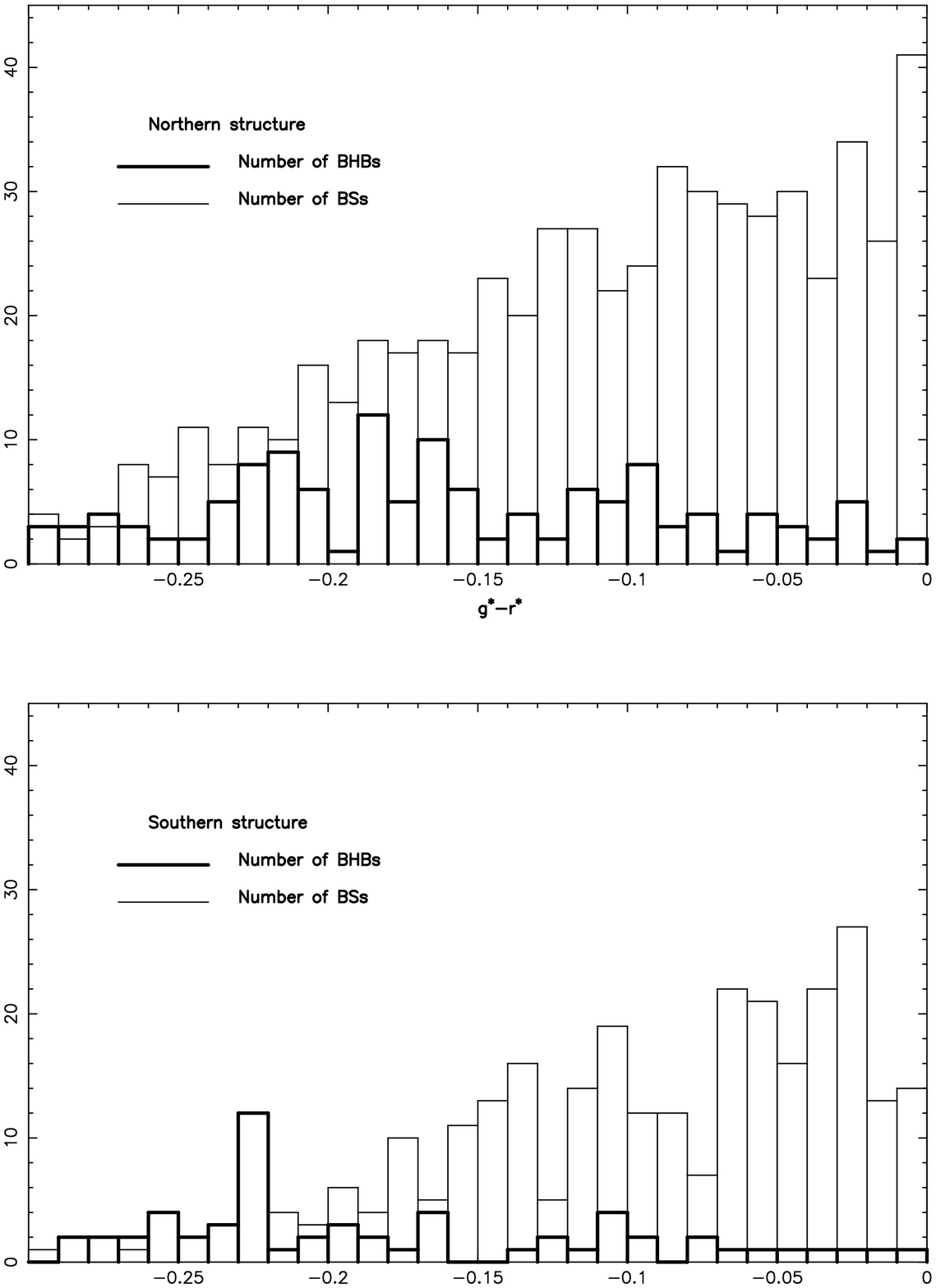}
\plotone{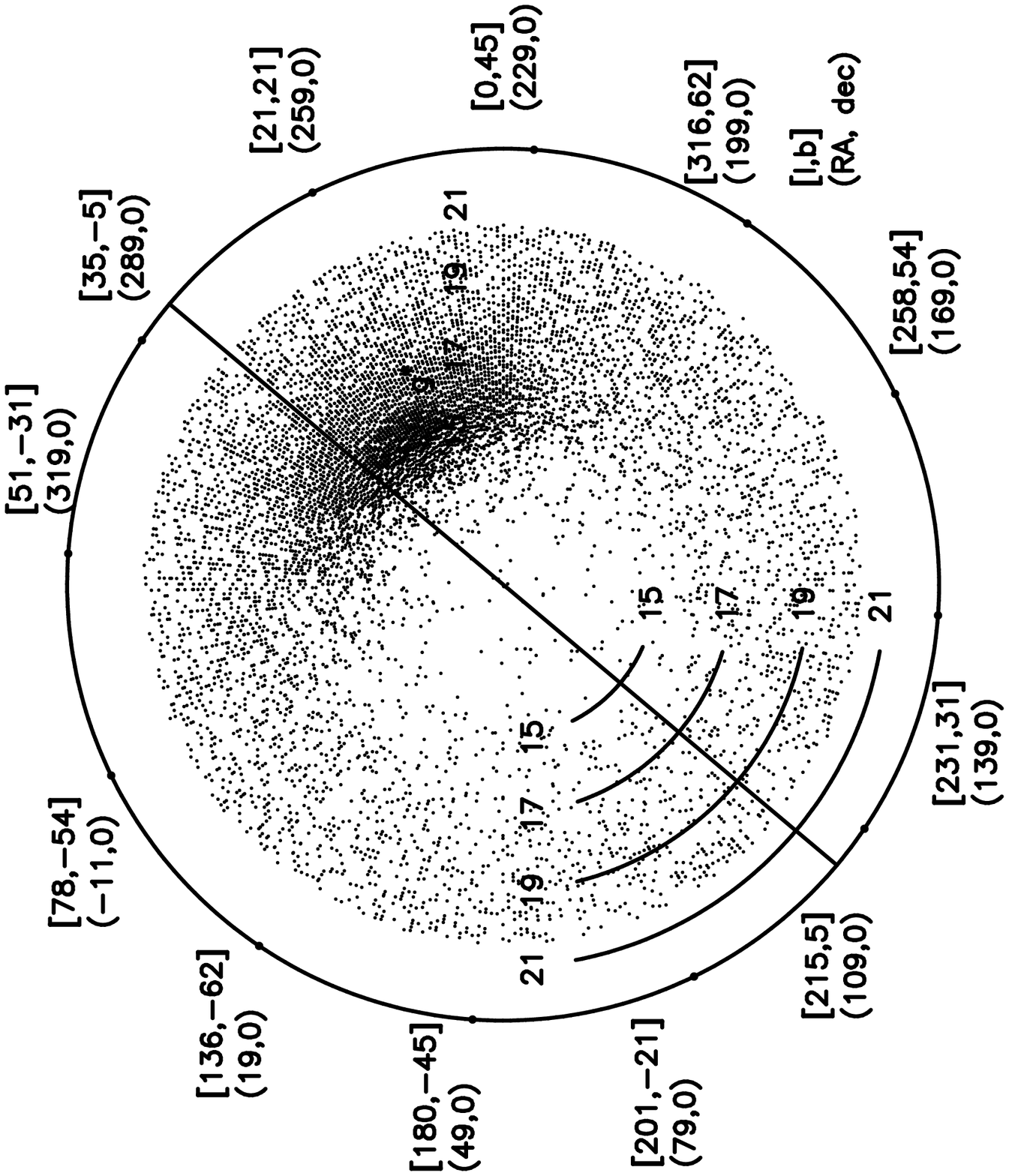}
\plotone{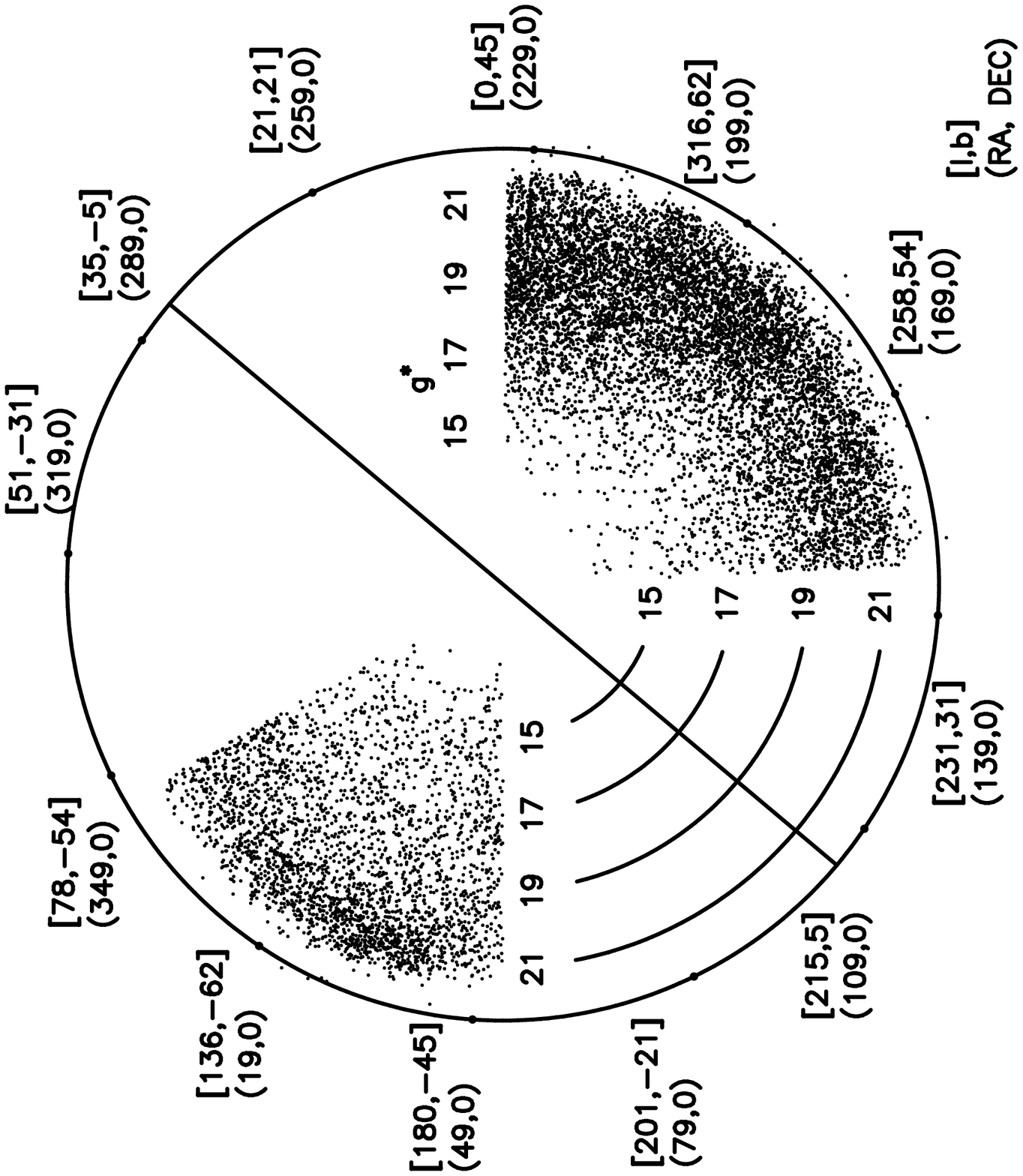}
\plotone{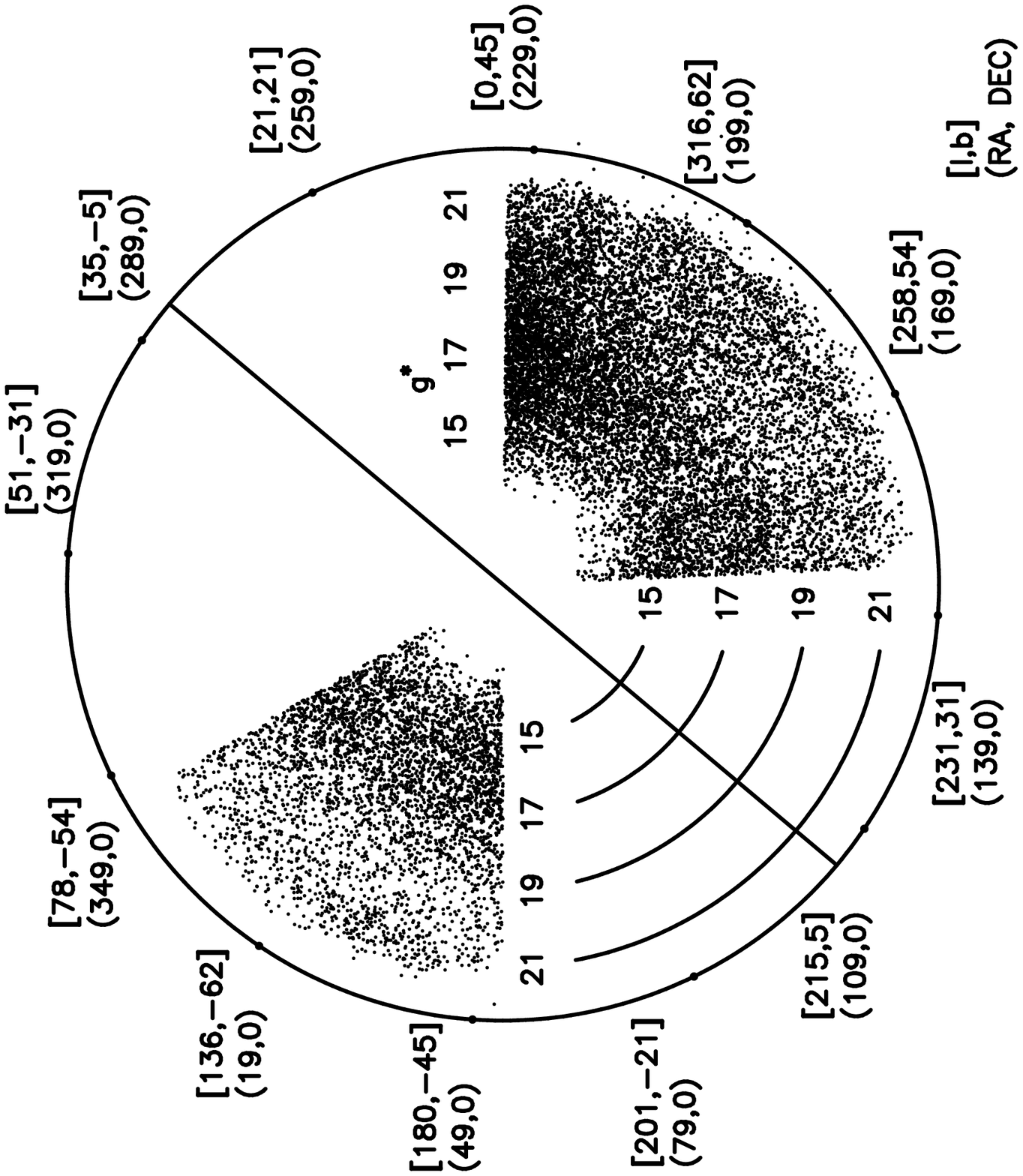}
\plotone{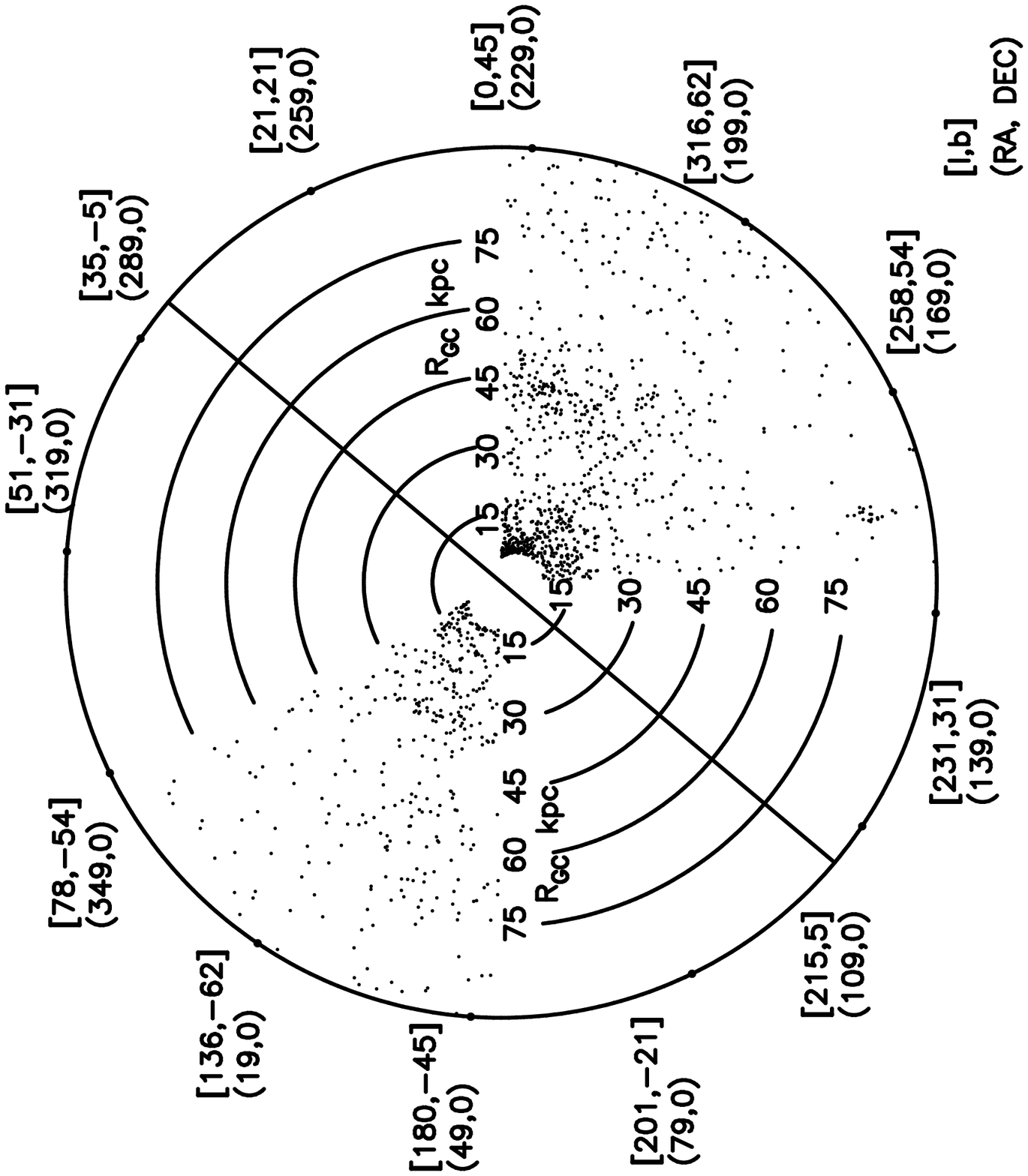}
\plotone{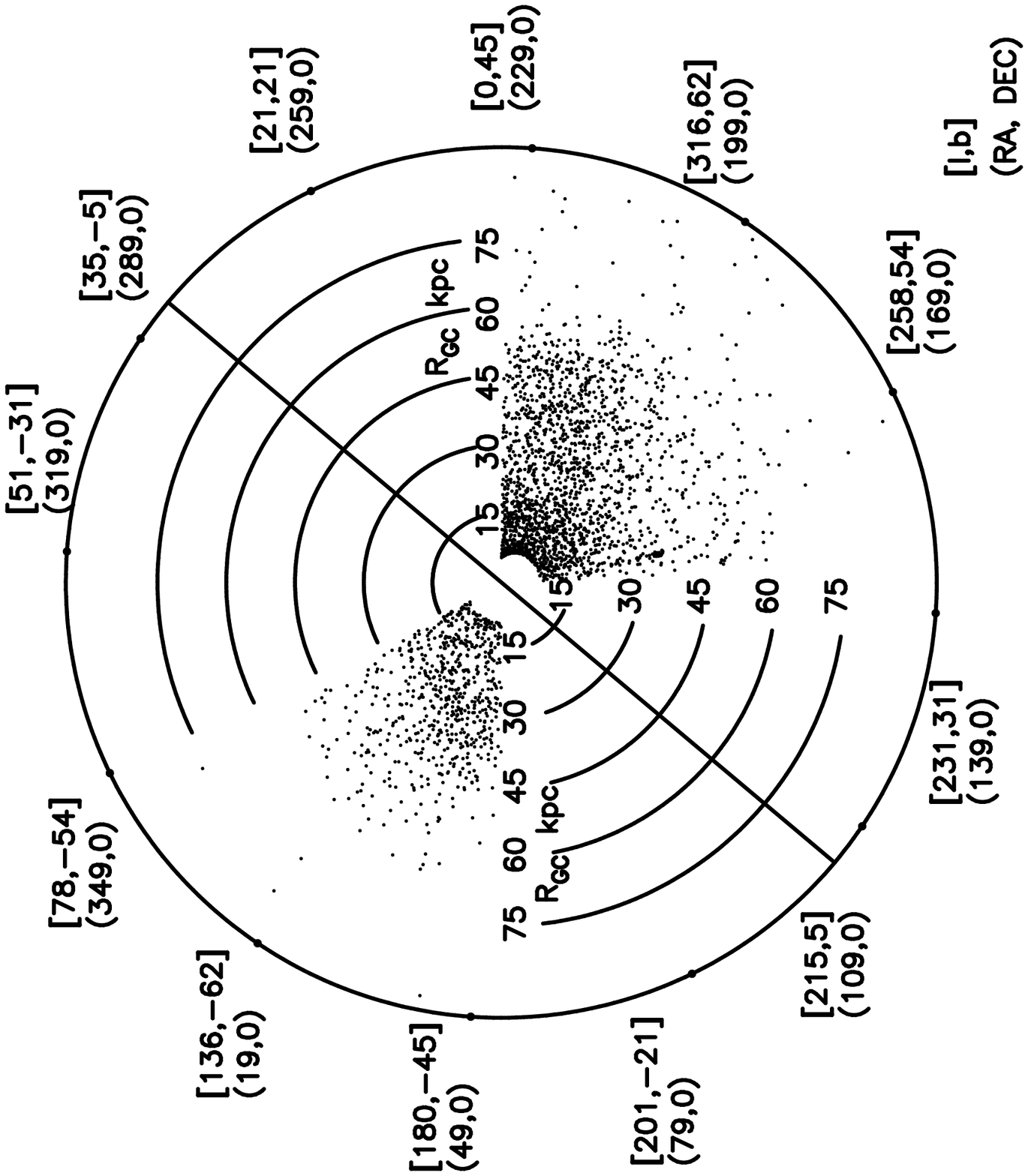}
\plotone{yfig20.ps}
\end{document}